\documentclass{emulateapj}
\usepackage{amsmath}
\usepackage{subfigure}
\usepackage{color}

\usepackage{mathrsfs}
\usepackage{graphicx}
\usepackage{cases}
\usepackage{txfonts}
\usepackage{epsfig}
\usepackage{multirow}
\usepackage{dcolumn}
\usepackage{bm}
\usepackage{amssymb}
\usepackage{multirow}
\usepackage{enumitem}
\usepackage{enumerate}
\usepackage[colorlinks,linkcolor=blue,citecolor=blue,urlcolor=blue]{hyperref}
\graphicspath{{figure/}}
\usepackage{lineno}

\def\prd{Phys. Rev. D}

\def\prl{Phys. Rev. Lett.}
\def\prx{Phys. Rev. X}
\def\plb{Phys. Lett. B}
\def\jcap{JCAP}
\def\apj{Astrophys. J.}
\def\apjl{Astrophys. J. Lett.}
\def\araa{Annu. Rev. Astron. Astrophys.}
\def\mnras{Mon. Not. Roy. Astron. Soc.}
\def\apjs{Astrophys. J. Suppl. Ser.}
\def\aanda{Astron. Astrophys.}

\def\cqg{Class. Quant. Grav.}

\def\ijmpd{Int. J. Mod. Phys. D}

\def \nat{Nature}
\def\lrr{Living Rev. Relativity}

\def\be{\begin{equation}}
\def\ee{\end{equation}}

\def\bea{\begin{eqnarray}}
\def\eea{\end{eqnarray}}
\newcommand{\bes}{\begin{equation*}}
\newcommand{\ees}{\end{equation*}}
\newcommand{\beqa}{\begin{eqnarray}}
\newcommand{\eeqa}{\end{eqnarray}}

\newcommand{\lsim}{\mathrel{\hbox{\rlap{\lower.55ex\hbox{$\sim$}} \kern-.3em \raise.4ex \hbox{$<$}}}}
\newcommand{\gsim}{\mathrel{\hbox{\rlap{\lower.55ex\hbox{$\sim$}} \kern-.3em \raise.4ex \hbox{$>$}}}}

\begin{document}

\title{Multi-messenger Detection Rates and distributions of Binary Neutron Star Mergers and Their Cosmological Implications}

\author{
Jiming Yu, $^{1,2\,\dagger}$, 
Haoran Song$^{3}$, 
Shunke Ai$^{4}$, 
He Gao$^{3}$,
Fayin Wang$^{5,6}$,
Yu wang$^{1,2}$,
Youjun Lu$^{7,8}$,
Wenjuan Fang$^{1,2}$,
and Wen Zhao$^{1,2\,\ddagger}$
\\
$^{1}$\,CAS Key Laboratory for Researches in Galaxies and Cosmology, Department of Astronomy, University of Science and Technology of China, Chinese Academy of Sciences, Hefei, Anhui 230026, China; $^\dagger$\,yjm8012@mail.ustc.edu.cn, $^\ddagger $\,wzhao7@ustc.edu.cn \\
$^{2}$\,School of Astronomy and Space Science, University of Science and Technology of China, Hefei, 230026, China \\
$^{3}$\,Department of Astronomy, Beijing Normal University, Beijing 100875, China \\
$^{4}$Department of Physics and Astronomy, University of Nevada Las Vegas, Las Vegas, NV 89154, USA \\
$^{5}$\,School of Astronomy and Space Science, Nanjing University, Nanjing 210093, China\\
$^{6}$\,Key Laboratory of Modern Astronomy and Astrophysics (Nanjing University), Ministry of Education, China\\
$^{7}$\,CAS Key Laboratory for Computational Astrophysics, National Astronomical Observatories, Chinese Academy of Sciences, Beijing, 100101,
China \\
$^{8}$\,School of Astronomy and Space Science University of Chinese Academy of Sciences, Beijing, 100049, China
}

\begin{abstract}
{

The gravitational-wave (GW) events, produced by the coalescence of binary neutron-stars (BNS), can be treated as the standard sirens to probe the expansion history of the Universe, if their redshifts could be determined from the electromagnetic observations. For the high-redshift ($z\gtrsim 0.1$) events, the short $\gamma$-ray bursts (sGRBs) and the afterglows are always considered as the primary electromagnetic counterparts. In this paper, by investigating various models of sGRBs and afterglows, we discuss the rates and distributions of BNS mergers' multi-messenger observations with GW detectors in second-generation (2G), 2.5G, 3G era with the detectable sGRBs and the afterglows. For instance, for Cosmic Explorer GW detector, the rate is about (300-3500) per year with GECAM-like detector for $\gamma$-ray emissions and LSST/WFST detector for optical afterglows. In addition, we find these events have the redshifts $z\lesssim 2$ and the inclination angles $\iota\lesssim 20^{\circ}$. These results justify the rough estimation in previous works. Considering these events as standard sirens to constrain the  equation-of-state parameters of dark energy $w_{0}$ and $w_{a}$, we obtain the potential constraints of $\Delta w_{0}\simeq 0.02-0.05$ and $\Delta w_{a}\simeq 0.1-0.4$. }

\end{abstract}

\section{Introduction}

\label{intro}

The discoveries of dozens of gravitational-wave (GW) signals produced by the inspiral and merger of compact binary systems \citep{abbott2016a,abbott2016b, abbott2016c, abbott2016d, abbott2017a, abbott2017b, abbott2017c, abbott2017d, abbott2019a, abbott2020a, abbott2020b, abbott2020c, abbott2020d} mark the opening of the era of GW astronomy. Since one can measure the luminosity distance of the GW source without any relying on a cosmic distance ladder \citep{schutz1986}, if the source's redshift can be measured independently, this kind of GW events can be treated as standard sirens to measure various cosmology parameters \citep{holz2005, sath2010, zhao2011, yan2020, wang2020}. On 2017 August 17, a GW event (GW170817) produced by a binary neutron star (BNS) system, together with a $\gamma$-ray burst (GRB 170817A) are observed \citep{abbott2017a, abbott2017d, goldstein2017, savchenko2017}. With the identification of their host galaxy, NGC 4993 \citep{arcavi2017, coulter2017, soares2017, tanvir2017, valenti2017}, advanced LIGO and Virgo collaborations (LVC) gave the first constraint on Hubble constant from standard sirens, $H_{0}=70^{+12.0}_{-8.0}\ \mathrm{km}\ \mathrm{s}^{-1}\ \mathrm{Mpc}^{-1}$ with 68.3\% confidence level \citep{ligo2017}. Many recent works also discussed the Hubble constant measurement through standard sirens \citep{chen2017, fishbach2019, soares2019, mortlock2019, gray2020, yu2020}. \par
Accurate measurement of luminosity distance $d_{L}$ is crucial to standard sirens. Building more or more advanced detectors will improve the $d_{L}$ measurement of the GW detection. At the same time, this will increase the redshift detection limit of GW signals, so GW standard sirens can be used to study the evolution of the high-redshift Universe. In the near future, KAGRA \citep{abbott2018} and LIGO-India \citep{unni2013}, will be built. Together with two advanced LIGO (aLIGO) and Virgo, there will be a network of five second-generation (2G) ground-based laser interferometer GW detectors. In \cite{ligo2016}, a modest cost upgrade of aLIGO, named LIGO A+, is proposed. Looking forward, two 3G detectors, including Einstein Telescope (ET) \citep{punturo2010, abernathy2011} and Cosmic Explorer (CE) \citep{abbott2017f, dwyer2015} are also under consideration.\par
The measurement of redshift is also an important task to standard sirens. One of the main methods is by observing their electromagnetic (EM) counterparts. The mergers of BNS and neutron star-black holes (NSBHs) are believed to create short GRBs \citep{paczynski1986, eichler1989, narayan1992, rosswog2013} and kilonovae \citep{li1998, rosswog2005, tanvir2013, metzger2017}. From the afterglow of GRBs or kilonovae, one can measure the GW source's redshift. Usually, the afterglows of GRB are with a much narrower observation angle but a larger redshift range \citep{berger2014}. Therefore, the observations of afterglows are more important for the standard sirens with high redshift. \par
In this work, we assume the jet profile obtained from the GRB170817A is quasi-universal for all BNS mergers and perform Monte Carlo simulations to estimate the magnitudes of their GW signals, GRB counterparts and afterglows. We select the samples that can be triggered by both GW interferometers and EM telescopes, and treat them as standard sirens. We get the standard sirens' rates and their distributions of redshifts and inclination angles. Moreover, we use these samples of standard sirens, combined with the method mentioned in \cite{sath2010} and \cite{zhao2011}, to discuss the implications for dark energy parameters constraining by 3G GW interferometers. \par

This paper is organized as follows. In Sec. \ref{samples}, we show the BNS samples used in this work. We introduce the detector response and the Fisher information matrix in Sec. \ref{fishermatrix}. The jet profile of GRB170817A, the observation abilities of $\gamma$-ray detectors and optical telescopes we used are mentioned in Sec. \ref{emcounterpart}. {\color{black}We estimate the detectability of BNS mergers' GW signals, GRB counterparts, afterglows, and discuss the implications of dark energy parameters constraints in Sec. \ref{multi-messenger}.} As a supplement, we discuss several alternative models for the EM counterparts in Sec. \ref{discussion}. In the end, we summarize our results in Sec. \ref{conclusion}.\par

Throughout the paper, we choose the unit with $c=G=1$, where $G$ is the Newtonian gravitational constant and $c$ is the speed of light in vacuum. We adopt the standard $\Lambda$CDM model with following parameters $H_{0}=67.8\mathrm{km}\ \mathrm{s}^{-1}\mathrm{Mpc}^{-1}$, $\Omega_{m}=0.308$, $\Omega_{\Lambda}=0.682$ \citep{planck2016}.

%
%

%

%
%
\section{BNS samples}
\label{samples}
The event rate of BNS mergers with redshift $z$ could be estimated as
\be
	N(z)dz=\frac{R_{\mathrm{BNS mergers},0}\times f(z)}{1+z}\frac{dV(z)}{dz}dz,
\label{nz}
\ee
where $R_{\mathrm{BNS mergers},0}$ is the local BNS merger rate and we set it as {\color{black}$R_{\mathrm{BNS mergers}, 0}=80-810\ \mathrm{Gpc}^{-3}\mathrm{yr}^{-1}$ in this work \citep{abbott2020e}}. $f(z)$ is the dimensionless redshift distribution function and $dV(z)/dz$ is the differential comoving volume, which is given by 
\be
	\frac{dV(z)}{dz}=\frac{4\pi c}{\mathcal{H}(z)}\chi^{2}(z),
\ee
where $\chi(z)$ is the comoving distant, $\mathcal{H}(z)$ is the conformal Hubble parameter with redshift.\par
The dimensionless redshift distribution function $f(z)$ depends on the initial distribution of the BNS system and their delay time from generation to merger. Here, we first assume the initial distribution follows the star formation rate (SFR). We adopt the model derived by  \cite{yuksel2008}:
\be
	\mathrm{SFR}(z)\propto\left [(1+z)^{3.4\eta}+\left (\frac{1+z}{5000}\right )^{-0.3\eta}+\left (\frac{1+z}{9}\right )^{-3.5\eta}\right ]^{1/\eta},
\label{sfr}
\ee
with $\eta=-10$ and in units of $M_{\odot}\ \mathrm{Gpc}^{-3}\mathrm{yr}^{-1}$. For delay time, we use the log-normal distribution model (\cite{wand2015}). In this module, the delay time $\tau$ has the distribution
\be
	P(\tau)d\tau=\frac{1}{\sqrt{2\pi}\tau\sigma}\exp\left [-\frac{\left (\ln\tau-\ln t_{d}\right )^{2}}{2\sigma^{2}}\right ]d\tau,
\label{tau}
\ee
with $t_{d}=2.9\ \mathrm{Gyr}$ and $\sigma=0.2$. 
The distribution of BNS merger samples can be integrated from Eq. (\ref{sfr}) and Eq. (\ref{tau})
\be
	f(z(t_{0}))=\int \mathrm{SFR}(z(t_{0}-\tau))P(\tau)d\tau.
\label{fz}
\ee
 The range of the initial distribution is from 0 to 7, with the same upper limit with \cite{yuksel2008}. So, we can get the BNS merger samples with the redshift distribution of  Eq. (\ref{nz}). Their distribution is isotropic. As the component masses $m_{1}$ and $m_{2}$, we choose a normal distribution $N~(\mu=1.32M_{\odot},\ \sigma=0.11M_{\odot})$ for simulation, which following the observationally deviation distribution of galactic BNS system \citep{kizi2013}. We denote the inclination angle between the binary’s orbital angular momentum and the line of sight as $\iota$. The distribution of $\iota$ is proportional to $\sin\iota$. All of the BNS mergers in our simulation are non-spinning systems.\par

\section{GW detection}
\label{fishermatrix}

In this work, we mainly consider the network with $N_{d}$ GW detectors and denote their spatial locations as $\mathbf{r}_{I}$ with $I=1,2,\cdots,N_{d}$. Here, we make an approximation that each detector has spatial size much smaller than the GW wavelength. It is worth noting that for 3G detectors, this approximation would introduce significant biases in localization \citep{essick2017}. Therefore, the frequency-dependent responses should be considered in the real localization. In the Fig. 5 of \cite{essick2017}, the results of localization with/without this approximation are showed. Beyond the bias between them, they have similar errors. Since we concentrate on the error estimate in this work, this bias is unimportant and has little impact on our results. 
For the $I$-th detector, the response to an incoming GW signal traveling in direction $\mathbf{n}$ could be written as a linear combination of two wave polarizations, 
\begin{equation}
	d_{I}(t)=F^{+}_{I}h_{\times}(t)+F^{\times}_{I}h_{\times}(t).
\end{equation}	
The I$_{th}$ detector's beam-pattern function is decided by the source's right ascension (RA) $\alpha$, declination (DEC) $\delta$, the polarization angle $\psi$, the position and the orientation of interferometers's arms.  In this paper, we consider the 2G interferometers LIGO (Livingston), LIGO (Handford), Virgo, KAGRA, LIGO-India and 3G interferometers ET, CE and an assumed CE-type detector. Our pervious paper, \cite{yu2020}, listed the parameters of these interferometers, which are also mentioned in \cite{jaranowski1998}, \cite{blair2015} and \cite{vitale2017}. For LIGO (Livingston), LIGO (Handford), Virgo, KAGRA, LIGO-India, we use the noise curve of the designed level for advanced LIGO (aLIGO) and A+ \citep{ligo2016}. For CE in the U.S. and the assumed CE-type detector in Australia, we use the proposed noise curve in \cite{abbott2017f} and \cite{dwyer2015}. And for ET, we consider the proposed ET-D project \citep{punturo2010, abernathy2011}. In this paper, we denote the network of LIGO (Livingston), LIGO (Handford) and Virgo with aLIGO-type of noise curve as LHV, and the network of five 2G interferometers as LHVIK. LHV A+ and LHVIK A+ are used to represent the networks with the noise curve of A+ type. CEET represents the network of CE in U.S. and ET in Europe and CE2ET is the network of two CE-type interferometers in U.S. and Australia, and one ET in Europe. \par
We adopt the restricted post-Newtonian (PN) approximation of the waveform for the non-spinning systems with only the waveforms in inspiralling stage \citep{sath2009, cutler1993, blan2002, blan2002b, blan2004a, blan2004b, blan2005, damour2001, erratum2005, itoh2001, itoh2003, itoh2004a, itoh2004b} in this paper. The terms $h_{+}$ and $h_{\times}$ are depended on the mass ratio $\eta\equiv m_{1}m_{2}/(m_{1}+m_{2})^{2}$, the chirp mass $\mathcal{M}_{c}\equiv(m_{1}+m_{2})\eta^{3/5}$, the $d_{L}$, the $\iota$, the merging time $t_c$ and the merging phase $\psi_{c}$. Therefore, for a BNS merger, the response of interferometer depends on nine parameters, $\bm{\theta}=\{\alpha,\delta,\psi,\iota,M,\eta,t_{c},\phi_{c},\log(d_{L})\}$. Employing the nine-parameter Fisher matrix $\Gamma_{ij}$ and marginalising over the over parameters, we derive the covariance matrix $(\Gamma^{-1})_{ij}$ for the nine parameters \citep{wen2010}. The signal-to-noise ratio (SNR) for the GW signal can also be derived from the interferometer's response (See \cite{zhao2018} for details of Fisher matrix and SNR). In this work, we choose SNR$>12$ as GW signal's threshold. Note that, in the realistic observation and analysis, various selection effects might induce the bias for the sources' parameters \citep{chen2017, fishbach2019, mortlock2019, chen2020}, which is beyond the scope of the current paper with Fisher matrix analysis. We leave it as a future work. 
\section{EM counterpart detection}
\label{emcounterpart}
\subsection{$\gamma$-ray Brust detection}

\label{gamma-ray}

The observations of GW170817/GRB 170817A suggest it has a Gaussian-shaped jet profile \citep{zhang2002, troja2018, alexander2018, mooley2018, lazzati2018, ghirlanda2019}
\be
	E(\iota)=E_{0}\exp\left (-\frac{\iota^{2}}{2\iota_{c}^{2}}\right )
\label{jet} 
\ee
for $\iota \leq \iota_{w}$, where $E_{0}$ is the on-axis equivalent isotropic energy, $\iota_{c}$ is the characteristic angle of the core, $\iota_{w}$ is the truncating angle of the jet.  \cite{troja2018} give the constraints on the jet with $\iota_{c}= 0.057^{+0.025}_{-0.023}$, $\log_{10}E_{0}=52.73^{+1.3}_{-0.75}$, $\iota_{w}=0.62^{+0.65}_{-0.37}$. We assume all BNS mergers in our simulation has a relativistic jet with this profile. In order to convent this energy profile to the luminosity profile, we assume the burst duration $T_{90}\sim 2$ s \citep{abbott2017e} and the spectrum is flat with time during the burst, where $T_{90}$ is defined as the duration of the period which 90\% of the burst's energy is emitted. So, the $\gamma$-ray flux of each BNS mergers can be written as
\be
	F_{\gamma}=\frac{E_{0}\eta_{\gamma}}{4\pi D_{L}^{2} T_{90}}\exp\left (-\frac{\iota^{2}}{2\iota_{c}^{2}}\right )
\label{flux},
\ee
where $\eta_{\gamma}$ is the radiative efficiency and we adopt it as 0.1 for the bolometric energy flux in the 1 - 10$^4$ keV band. 
We assume the spectrum for all BNS-associated-GRBs follow the Band function 
\be
    N(E)= A\\
    \begin{cases}
        \left(\frac{E}{100\ \mathrm{keV}}\right )^{\alpha}\exp\left[-\frac{(2+\alpha)E}{E_{p}}\right],&E\leq\frac{(\alpha-\beta)E_{p}}{2+\alpha},\\
        \left(\frac{E}{100\ \mathrm{keV}}\right )^{\beta}\left[\frac{(\alpha-\beta)E_{p}}{(2+\alpha)100\ \mathrm{keV}}\right]^{\alpha-\beta}e^{\beta-\alpha},& E\geq\frac{(\alpha-\beta)E_{p}}{2+\alpha}.
    \end{cases}
\ee
Here $N(E)$ is in units of photons cm$^{-2}$ s$^{-1}$ keV$^{-1}$ and $E_{p}$ corresponds to the peak energy in the $\nu F_{\nu}$ spectra. We adopt the photon indices $\alpha$ and $\beta$ as, -1 and -2.3 \citep{preece2000} below and above the peak energy $E_{p}$ respectively. However, the bolometric isotropic luminosity $L$ and the peak energy for GRB 170817A do not follow the $E_p-L$ relation for short $\gamma$-ray bursts \citep{Zhang2018} and long $\gamma$-ray bursts \citep{yonetoku2004}, which can be written as
\be
    \frac{L}{10^{52}\ \mathrm{erg}\ \mathrm{s}^{-1}}=\left(2.34^{+2.59}_{-1.76}\right)\times10^{-5}\left[\frac{E_{p}(1+z)}{1\ \mathrm{keV}}\right]^{2.0^{\pm0.2}}.
\ee
There is a profile proposed by \cite{ioka2019} that the peak energy $E_{p}$ changes with $\iota$ with following relationship 
\be
	E_{p}(\iota)=E_{p,0}\times(1+\iota/\iota_{c})^{-0.4},
\ee
within the Gaussian structure jet framework, where $E_{p,0}$ is the peak energy that satisfies the Yonetoku relation. This profile makes the GRB170817a observations consistent with previous GRB observation. \par
With the peak energy and the Band function, we can get the effective sensitivity limit for various $\gamma$-ray detectors. Here we adopt the same setting as \cite{song2019}, the sensitivity for Fermi-GBM is adopted as $\sim2\times10^{-7}\ \mathrm{erg}\ \mathrm{s}^{-1}\ \mathrm{cm}^{-2}$ in 50 keV to 300 keV \citep{meegan2009}, the sensitivity for GECAM is adopted as $\sim1\times10^{-7}\ \mathrm{erg}\ \mathrm{s}^{-1}\ \mathrm{cm}^{-2}$ in 50 keV to 300 keV \citep{zhang2018}, the sensitivity for Swift-BAT and SVOM-ECLAIRS is adopted as $\sim1.2\times10^{-8}\ \mathrm{erg}\ \mathrm{s}^{-1}\ \mathrm{cm}^{-2}$ in 15 keV to 150 keV \citep{gehrels2019, gotz2014}, and the sensitivity for {\color{black}Einstein Probe (EP)} is adopted as $\sim3\times10^{-9}\ \mathrm{erg}\ \mathrm{s}^{-1}\ \mathrm{cm}^{-2}$ in 0.5 keV to 4 keV \citep{yuan2018}. Another important factor affecting the detection ability of the $\gamma$-ray detector is the size of its field of view. For Fermi-GBM, its field of view (FOV) covers about 3/4 of the whole sky and for GECAM, the field of view is about 4$\pi$. For other three detectors, their fields of view are much smaller. The factors are $\sim$ 1/9 for Swift-BAT, $\sim$ 1/5 for SVOM-ECLAIRS \citep{chu2016} and $\sim$ 1/11 for EP compared with the whole sky.  Note that, these $\gamma$-ray detectors might have no overlapped observational time with future 3G GW detector networks. However, we expect the similar or more powerful detectors in 3G era. So, for simplification, throughout this paper we only consider these $\gamma$-ray detectors for illustration. 

\subsection{Afterglows of $\gamma$-ray burst}
\label{afterglow}
The measurements of GRBs' redshift relies on the identification of their host galaxies and further optical observations. And the afterglow usually plays a crucial role in the host galaxy identification. In this paper, we use the standard afterglow models \citep{sari1998} to estimate the afterglows on $R$ band. The spectra and light curve are determined by the inclination angle $\iota$, the half-opening angle $\theta_{j}$, the total kinetic energy $E_{j}=(1-\cos\theta_{j})E_{0}$, the number density of the interstellar medium (ISM) $n_{0}$, the magnetic field energy fraction $\epsilon_{B}$, the accelerated electron energy fraction $\epsilon_{e}$, the power-law index of shock-accelerated $p$, the luminosity distance $d_{L}$ and the time after merger in unit of days $t_{j}$.\par

{\color{black}With adopting the convention notion $Q=10^{x}Q_{,x}$,} for a adiabatic jet, the Lorentz factor is given by \citep{sari1998}
\be
	\gamma(t)=8.9(1+z)^{3/8}E_{j,51}^{1/8}n_{0}^{-1/8}\theta_{j, -1}^{-1/4}t_{,d}^{-3/8}.
\label{gamma_t}
\ee
For an on-axis observer, a phenomenon named jet break is expected when $\gamma$ drops below $\theta_{j}$ and the jet's material begin to spread sideways \citep{sari1998, mezsaros1999}. At the jet break time, there will be a break in the afterglow's light curve. From Eq. (\ref{gamma_t}), the jet break time $t_{j}$ is given by
\be
	t_{j}=0.82(1+z)E_{j, 51}^{1/3}n_{0}^{-1/3}\theta_{j, -1}^{2}\ \mathrm{days}.
\ee
At the jet-break time, there are two cases for flux density, $F_{\nu, j}$, the slow-cooling case ($\nu_{m}<\nu<\nu_{c}$) and the fasting cooling cases ($\nu_{c}<\nu$), where $\nu_{m}$ is the typical synchrotron frequency of the accelerated electrons with the minimum Lorentz and $\nu_{c}$ is the cooling frequency. In the slow-cooling case, $F_{\nu, j}\propto \nu^{-(p-1)/2}$, where we adopt $p=2.2$ in this paper (the same as \cite{nakar2002} and \cite{zou2007}). While in the fast-cooling case, $F_{\nu, j}\propto\nu^{-p/2}$ \citep{sari1998}. \par
For an on-axis observer, the light curve is divided into two power-law segments by the jet break time \citep{sari1998}. We denote the temporal decay index of the flux density $F_{\nu,0}(t)$ in the time as $\alpha_{1}$ and $\alpha_{2}$, which represent the time before and after $t_{j}$, respectively. The index $\alpha_{1}$ is $-(2-3p)/4$ in the fast cooling case and $-3(1-p)/4$ in the slow cooling case. When $t>t_{j}$, the on-axis observer can only observe $\sim\theta_{j}^{2}\gamma^{2}$ of the flux density in the isotropic fireball case. Since $\gamma(t)\propto t^{-3/8}$, we have $\alpha_{2}=\alpha_{1}+3/4$.\par
For a point source moving at angle $\iota$, the observed flux is \citep{granot2002}
\be
	F_{\nu}=\frac{L'_{\nu'}}{4\pi d_{A}^{2}}\left (\frac{\nu}{\nu'}\right ),
\ee
where $d_{A}$ is the angular distance, $L'_{\nu'}$ and $\nu'$ are the jet comoving frame spectral luminosity and frequency. For an off-axis observer $(t, \nu)$ at the angle $\iota$ and an on-axis observer $(t_{0},\nu_{0})$ at the angle 0, there are $t_{0}/t\approx\nu/\nu_{0}=(1-\beta)/(1-\beta\cos\iota)\equiv a$, where $\beta=\sqrt{1-1/\gamma^{2}}$. And we obtain
\be
	F_{\nu}(\iota, t)=a^{3}F_{\nu/a}(0, at).
\label{fnu}
\ee
Here we consider both fast and slow cooling cases, which are $\nu_{c}<\nu$ and $\nu_{m}<\nu<\nu_{c}$, respectively.

\section{Multi-messenger observation of GW and GRB}
\label{multi-messenger}

\subsection{Detection rates}
Since we find $z=3$ has exceeded the detection limit of the GRB detectors in the simulations, we set it as the upper redshift limit and sampled $10^7$ BNS mergers. However, this upper limit is too high for the 2G interferometers that the simulation became very inefficient with them. The upper detection limit with LHVIK A+ is about $z\sim0.2$, so for the 2G interferometers, we resample them with $z\in(0,0.3]$.\par 
There are $\sim3.4\times10^{5}$ BNS samples detected by ET and $5.1\times10^{6}$ detected by CE. For 2G GW detector networks LHV, LHVIK, LHV A+ and LHVIK A+, the numbers are $2.7\times10^{4}$, $5.7\times10^{4}$, $1.6\times10^{5}$ and $3.3\times10^{5}$, respectively, with samples' $z\in(0, 0.3]$. The observation ability of A+ type network will improve several times even with less detectors. About 10-12\% BNS mergers' GW signal observed by 2G GW detector networks could be triggered by $\gamma$-ray detector. For 3G detectors, the fractions are much lower because more events have high redshift.\par 
{\color{black}We convert these numbers into rates per year, after multiplying the total rate of BNS mergers. Integrating Eq. (\ref{nz}), in $z\in(0,0.3]$, the total rate is 1159-11736 yr$^{-1}$ and in $z\in(0-3]$, the total rate is $6.9\times10^{4}-7.0\times10^{5}$ yr$^{-1}$.} Due to the limited observation area of the $\gamma$-ray detectors, not all GRB events whose flux reaches the threshold count be observed. So the multi-messenger observation rates require a discount, FOV/$\Omega_{0}$, where $\Omega_{0}$ is the solid angle of whole sky. Table \ref{table2} lists the observation rates per year with this discount. For LHV, the rate is 0.042-0.425 per year with Swift-BAT and 0.072-0.731 per year with SVOM-ECLAIRS. For GECAM and Fermi-GBM, the rates are a few times larger due to their much larger observation areas. The result of EP is slightly worse than Swift-BAT due to its smaller observation area, although it has better sensitivity. {\color{black}After adding two LIGO-type detectors, the rates of LHVIK are about twice as large as LHV.} The future upgrade of aLIGO, A+ would increase these numbers by about five times. For 3G interferometers ET, their will be about $100\sim800$ multi-messenger observation rate with different $\gamma$-ray detectors if we adopt 810 Gpc$^{-3}$yr$^{-1}$ as the local merger rate. The rates are much larger even only with one single interferometer. {\color{black}For another 3G interferometer CE, the rates are a few times larger compared with ET, since it has a further observation range.} It is worth noting that we assume the GW and GRB detections are independent. However, in the real situation, with a triggered GRB, one can search weaker GW signals. So the numbers in Table \ref{table2} might be slightly smaller compared to the actual situation. \par

\begin{figure*}[htbp]
\centering
\subfigure[2G]{
	\includegraphics[width=8cm]{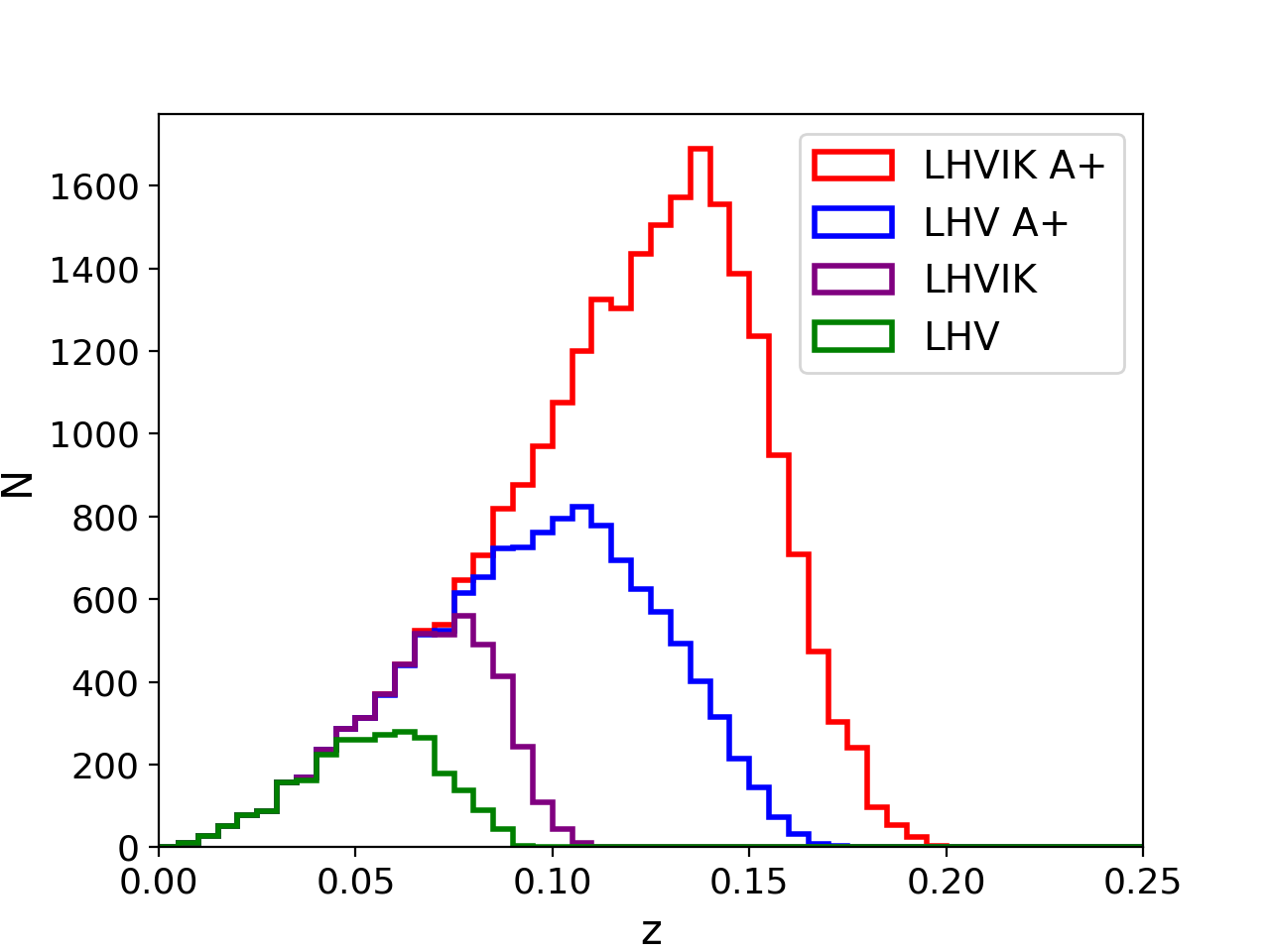}
}
\hspace{2pt}
\subfigure[3G]{
	\includegraphics[width=8cm]{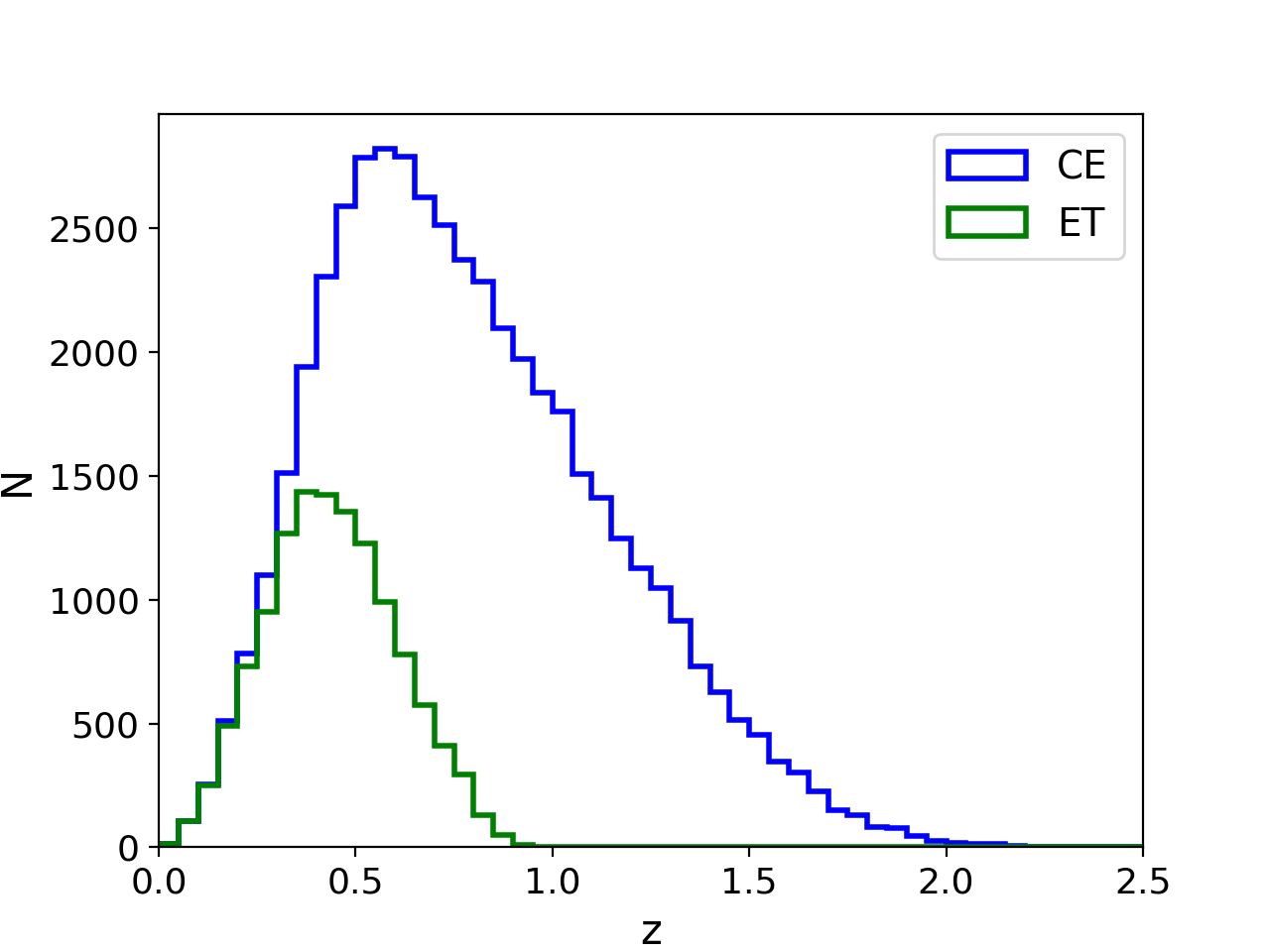}
}
\caption{The number distributions of samples that can be both triggered by GW interferometers and GECAM.}
\label{pdf_z}
\end{figure*}

We select the BNS samples that can be triggered by both GW interferometers and $\gamma$-ray detectors. We choose the result of GECAM as a representative and show their redshift distribution in Fig. \ref{pdf_z}. The redshift limits of LHV and LHVIK are around $z\sim0.1$ and the the detection limit of LHVIK A+ can reach $\sim0.2$ due to more sensitive detection capabilities. For 3G interferometers ET and CE, the redshift limits are $z\sim1.0$ and $z\sim2.5$, respectively.\par

\begin{figure}[htbp]
\centering
\includegraphics[width=8cm]{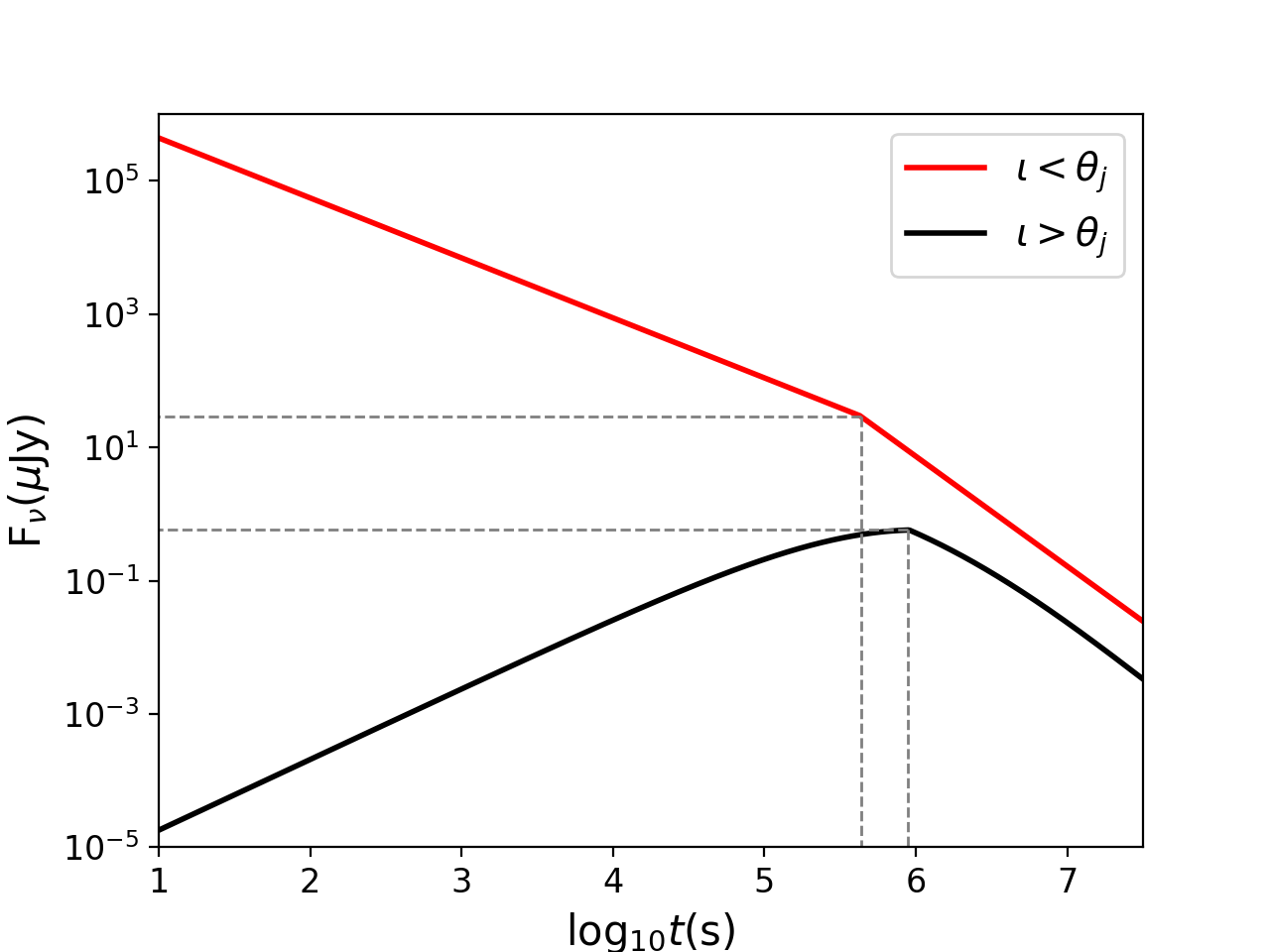}
\caption{This figure show the light curve of on-axis and off-axis afterglows. The red and black curves are the cases of on-axis afterglow with $\iota=5^{\circ}$ and off-axis afterglow with $\iota=15^{\circ}$, respectively. Other parameters in these two cases are same, with $E_{0}=10^{52.73}$ erg, $z=1$ and $\theta_{j}=10^{\circ}$.}
\label{afterglow_lc}
\end{figure}

As an important method for measuring redshift information, we also calculate the magnitude of GRB afterglows with the method mentioned in Subsection \ref{afterglow}. We adopt $n_{0}=1\ \mathrm{cm}^{-3}$, $p=2.2$, $\epsilon_{e}=0.1$, $\epsilon_{B}=0.01$, $p=2.2$, where they are the same as \cite{nakar2002} and \cite{zou2007}. For the half opening angle, we consider three cases, $\theta_{j}=5^{\circ},\ 10^{\circ},\ 15^{\circ}$. For off-axis samples, we calculate the afterglows' maximum $R$ band magnitude $M_{R}$ with different opening angles respectively. However, for on-axis GRB samples, their afterglows power-low decay with time and we cannot get their maximum $M_{R}$ from their light curve, as showed in Fig. \ref{afterglow_lc}, so we denote the $R$ band flux in the jet break time as $M_{R}$ in this case, as a comparison with the off-axis samples. \par

\begin{figure*}[htbp]
\centering
\subfigure[LHV]{
	\label{LHV_flux7_scatter}
	\includegraphics[width=8cm]{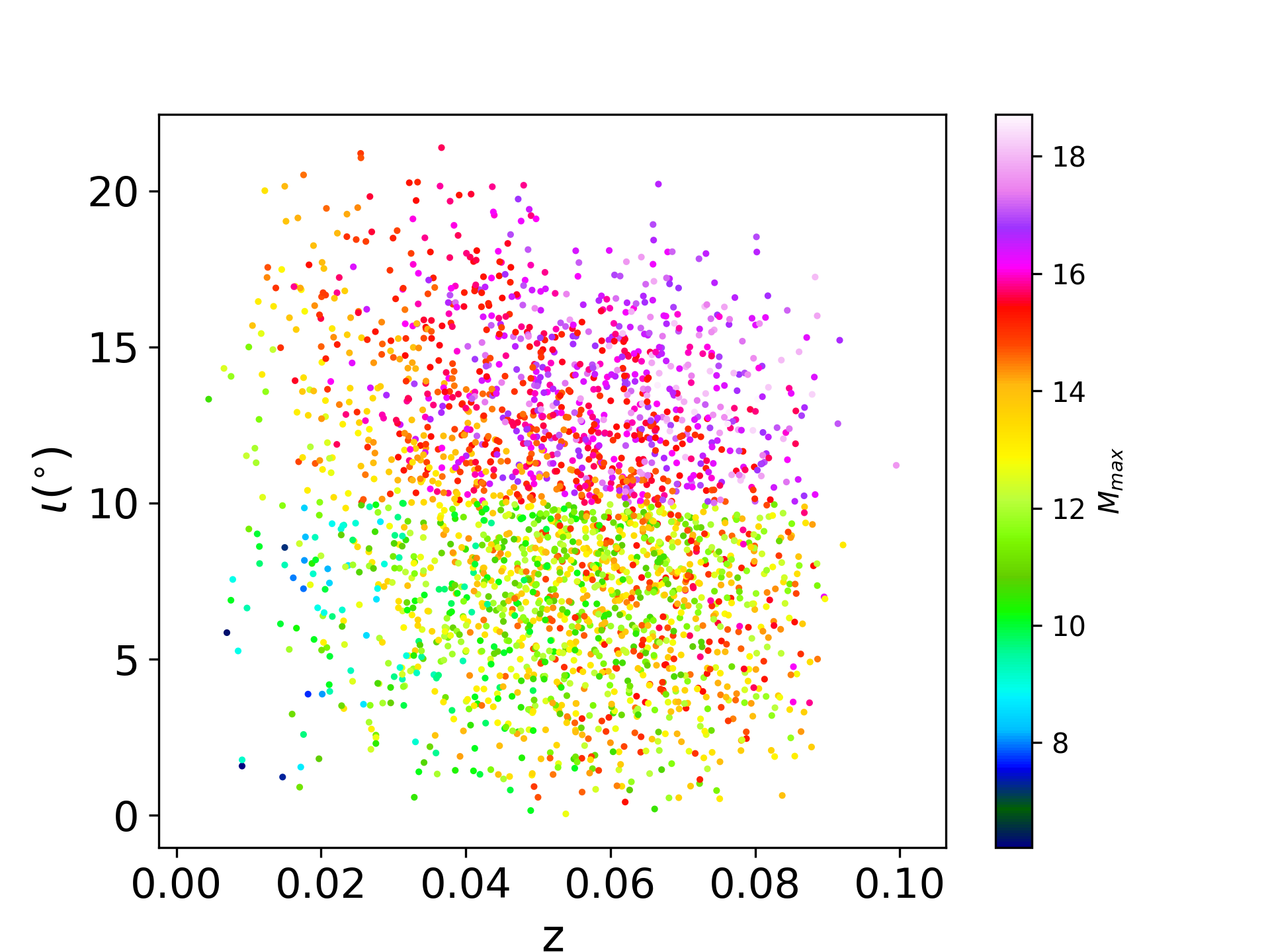}
}
\subfigure[LHVIK]{
	\label{LHVIK_flux7_sactter}
	\includegraphics[width=8cm]{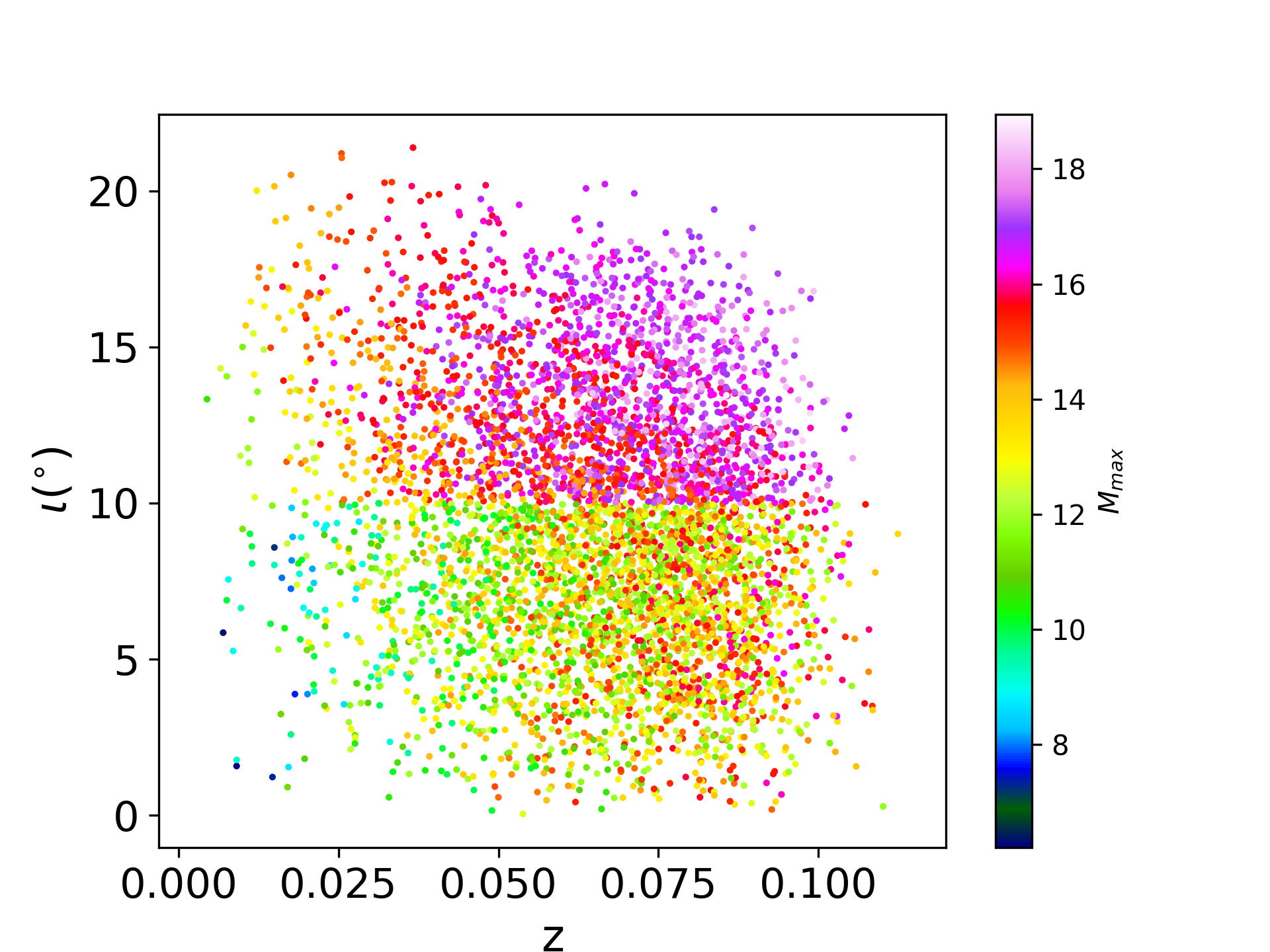}
}
\subfigure[LHV A+]{
	\label{LHV_A+_flux7_sactter}
	\includegraphics[width=8cm]{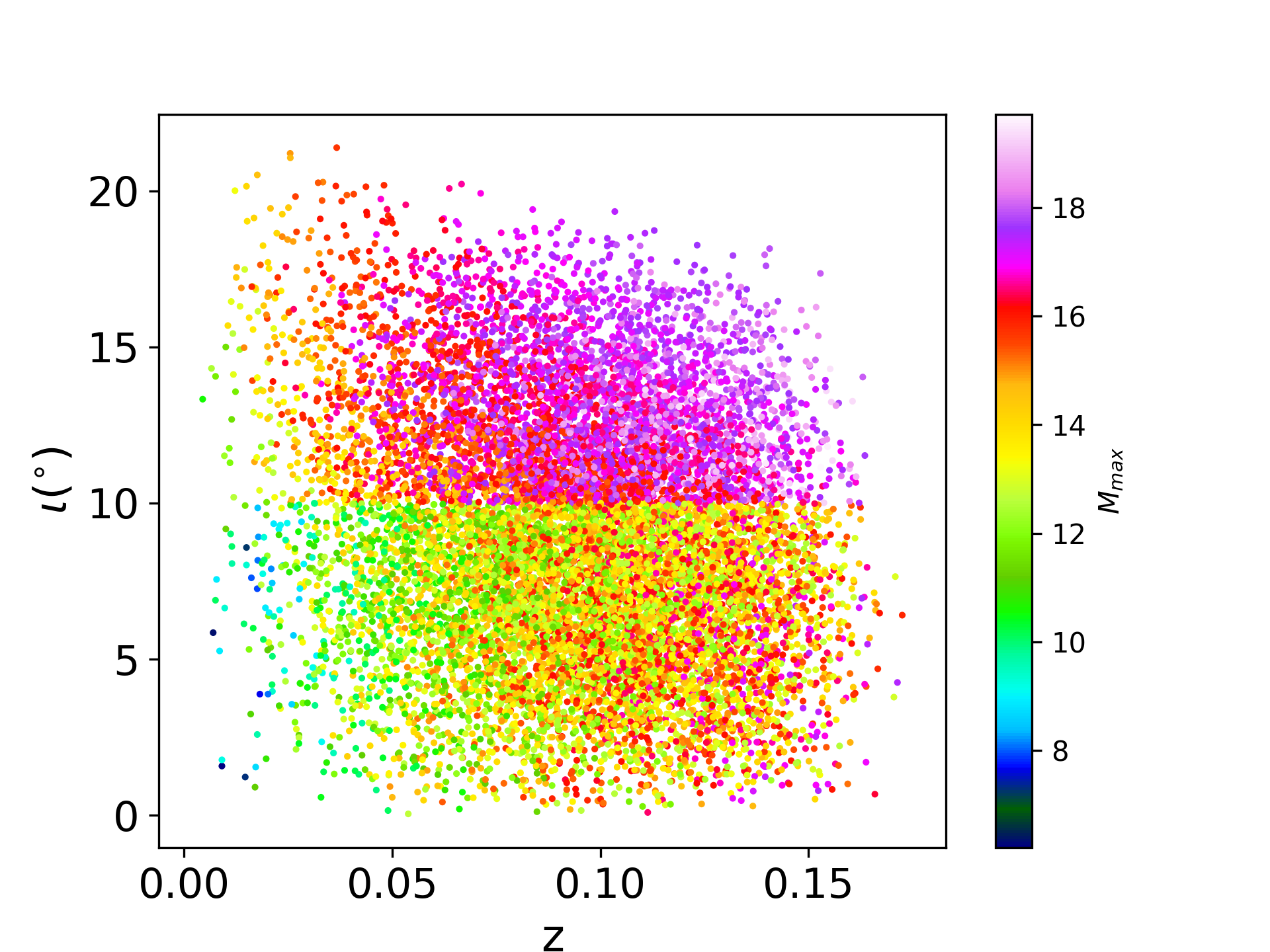}
}
\subfigure[LHVIK A+]{
	\label{LHVIK_A+_flux7_sactter}
	\includegraphics[width=8cm]{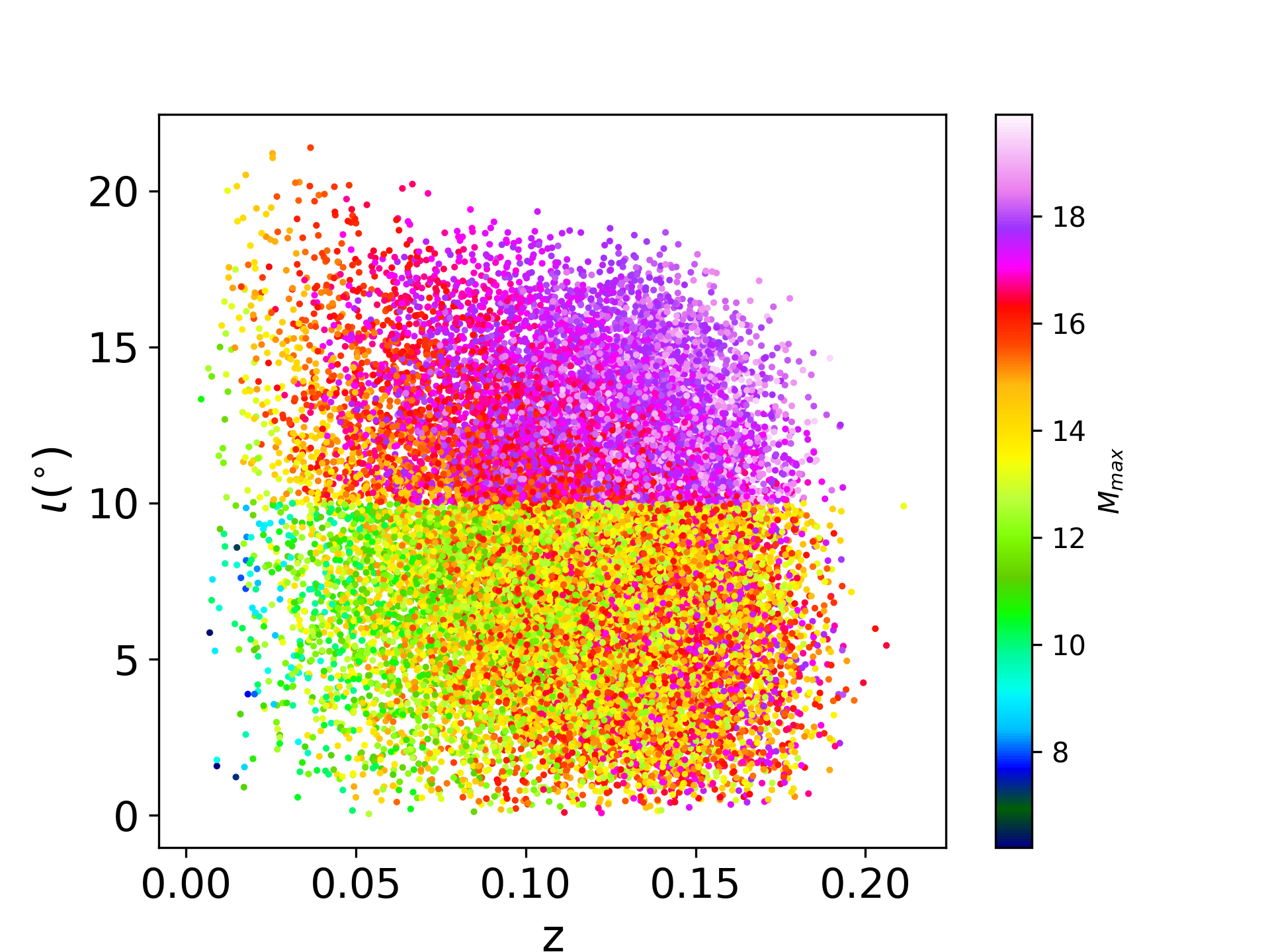}
}
\subfigure[ET]{
	\label{ETD_flux7_sactter}
	\includegraphics[width=8cm]{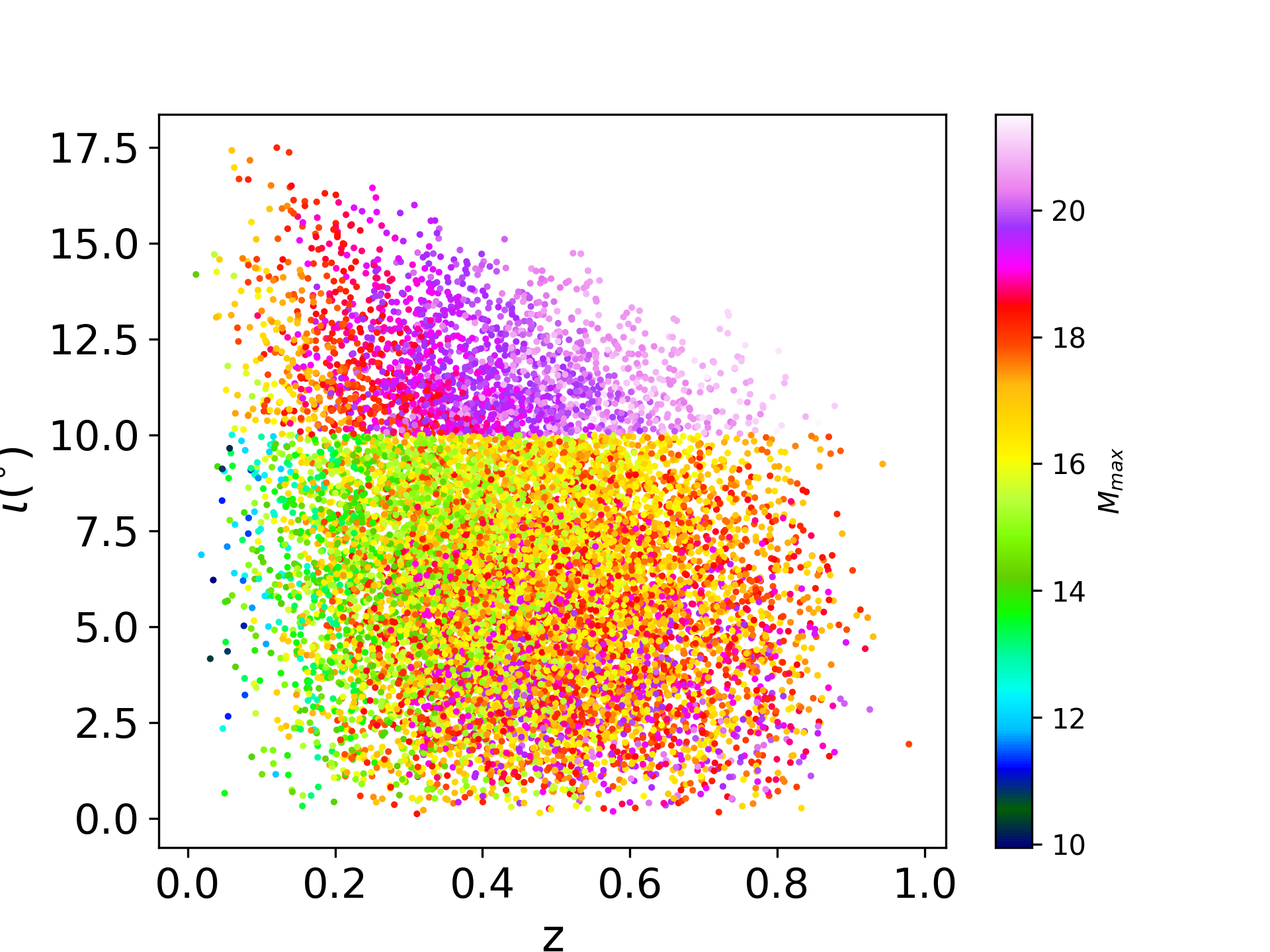}
}
\hspace{2pt}
\subfigure[CE]{
	\label{CE_flux7_sactter}
	\includegraphics[width=8cm]{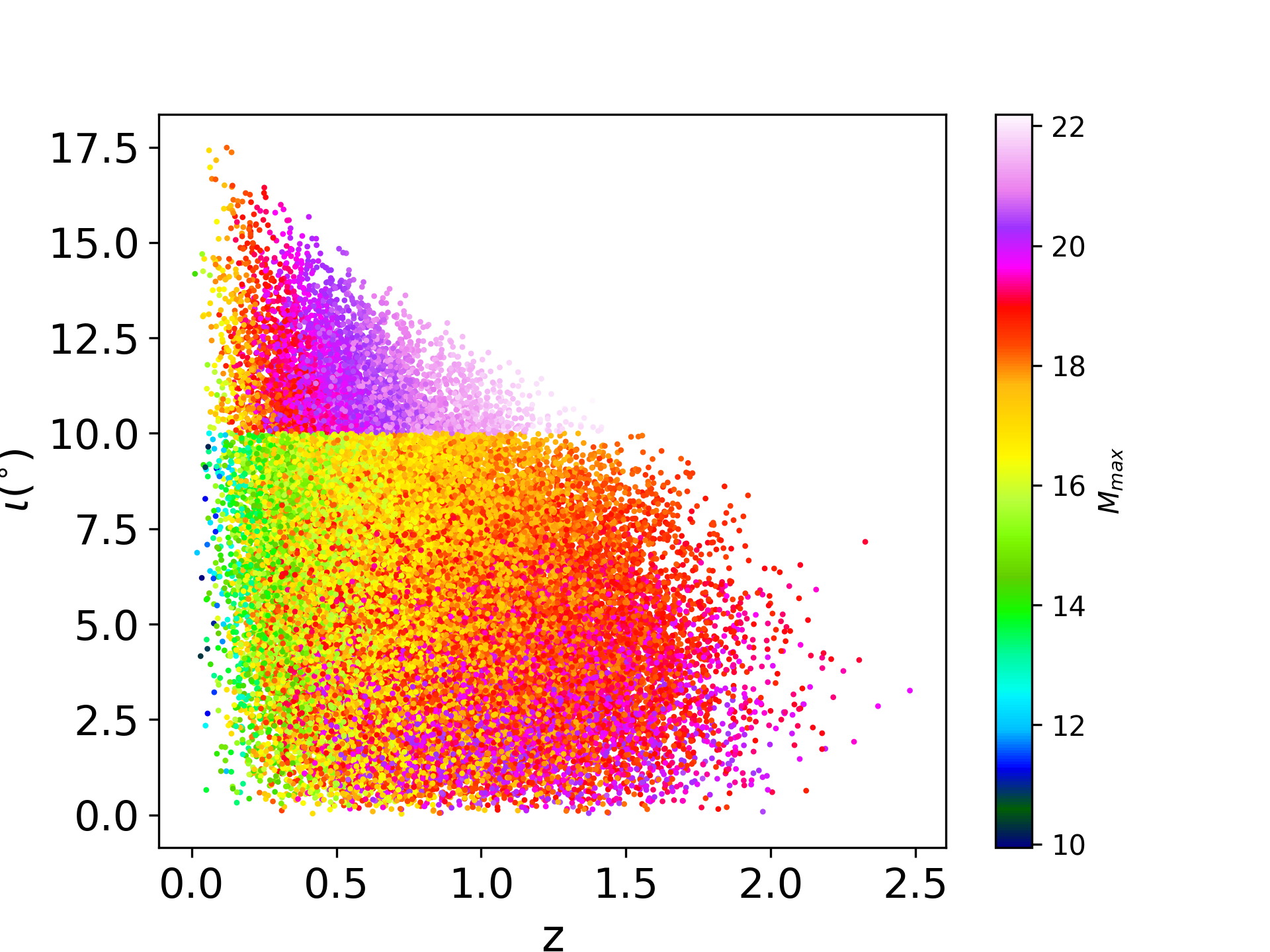}
}
\caption{These figures show the distributions of $\iota$, redshift of BNS samples and their afterglows flux, which can be triggered by GW detectors and GECAM. The upper and middle four panels with 2G GW detector networks LHV, LHVIK, LHV A+, LHVIK A+, show the results of $10^{7}$ samples with $z\in(0,0.3]$. The lower panels with ET and CE show the results of $10^{7}$ samples with $z\in(0,3]$. The colorbars show their R band magnitude of afterglows with $\theta_j=10^{\circ}$.}
\label{flux_scatter}
\end{figure*}

Using the case of GECAM and $\theta_{j}=10^{\circ}$ as an example, we show the distributions of standard sirens' redshift, inclination angle and the flux of afterglows in Fig. \ref{flux_scatter}.  From these figures, we can intuitively see the $\iota$ limit of multi-messenger observation. In very low redshift, there is a BNS sample with $\iota\sim22^{\circ}$. This sample's inclination angle and redshift are very similar with GW170817. For samples with $z=1$, the $\iota$ limit becomes to about $15^{\circ}$. \cite{zhao2011} and \cite{sath2010} have roughly estimated the distribution of GW sources on the order of magnitude, which is about 1000 per year with inclination angle $\iota<20^{\circ}$. These predictions are based on the $\sim$ several $\times 10^{5}$ BNS rate per year within the horizon of ET, and $\sim 10^{-3}$ of them will have GRBs toward us. These estimates meet well with our result. We use the improved measurement of BNS merger rate from GW observations and give stronger constraints on the multi-messenger event rates. In the next subsection, we will use their methods, combined with our results of simulations, to show the potential of determination of dark energy by the 3G GW interferometers ET and CE. The design specification of LSST for $5\sigma$ depths for point sources in the $r$ band is $\sim24.7$ \citep{ivezic2019}, which correspond to point sources and fiducial zenith observations. Due to the small inclination angles, all standard sirens have $M_{R}$ brighter it. For 2.5-meter Wide Field Survey Telescope (WFST), the sigle-visit depth with a 30 s exposure time is $r\sim22.8$. A small fraction of samples in the case of SVOM fainter than $r\sim22.8$. If we consider a 300 s exposure time, the sigle-visit depth of WFST will reach $r\sim24.1$ and all samples have $M_{R}$ brighter than it. In other two choices of $\theta_{j}$, their afterglows all reach the design depth of LSST and WFST. Therefore, we have assume all multi-messenger BNS events in our simulation have a detectable afterglow and we can measure their redshift from their afterglows.
\par
\begin{figure*}[htbp]
\centering
\subfigure[ET, $n_{0}=0.1\ \mathrm{cm}^{-3}$]{
	\includegraphics[width=8cm]{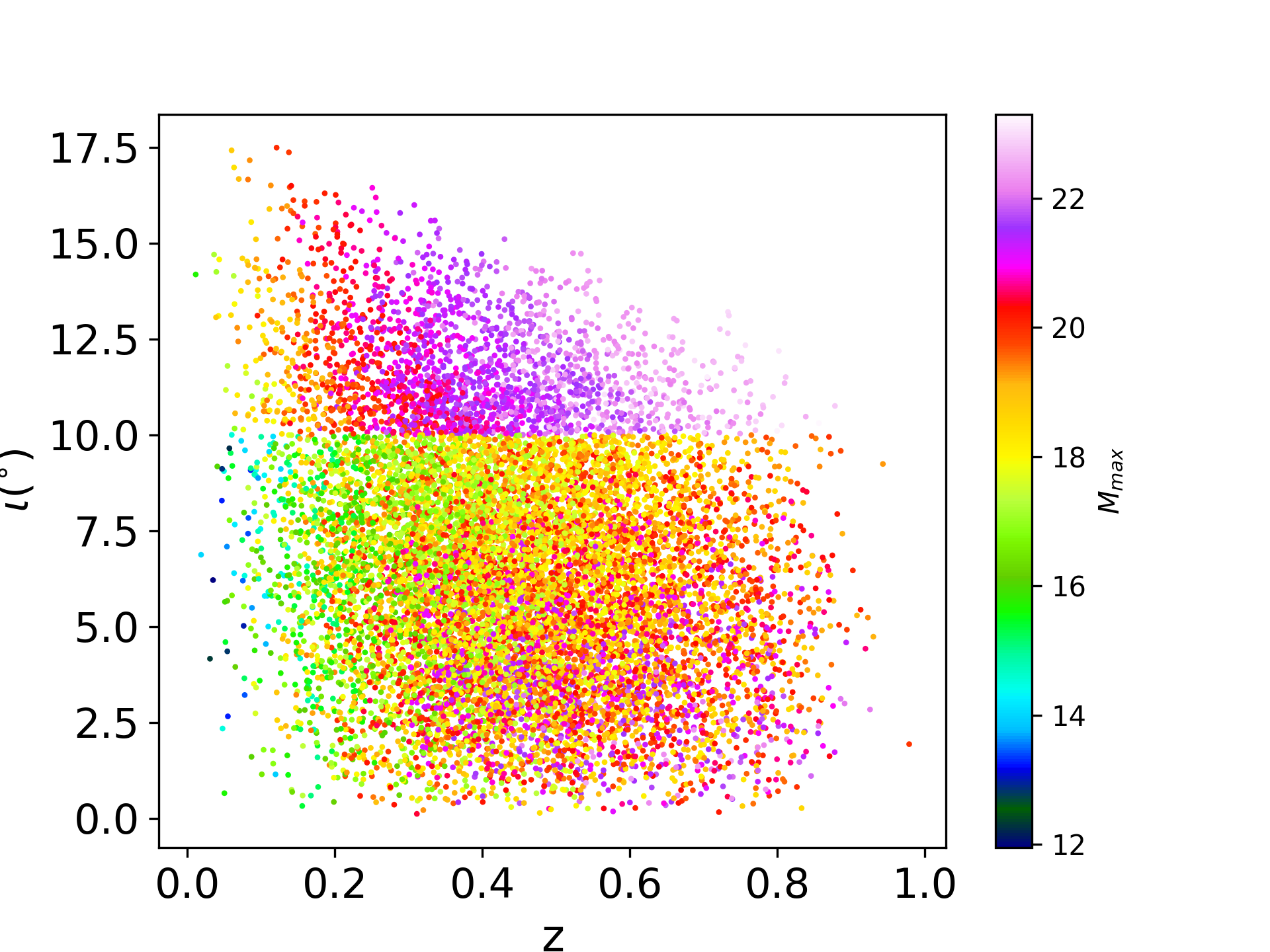}
}
\hspace{2pt}
\subfigure[ET, $n_{0}=0.01\ \mathrm{cm}^{-3}$]{
	\includegraphics[width=8cm]{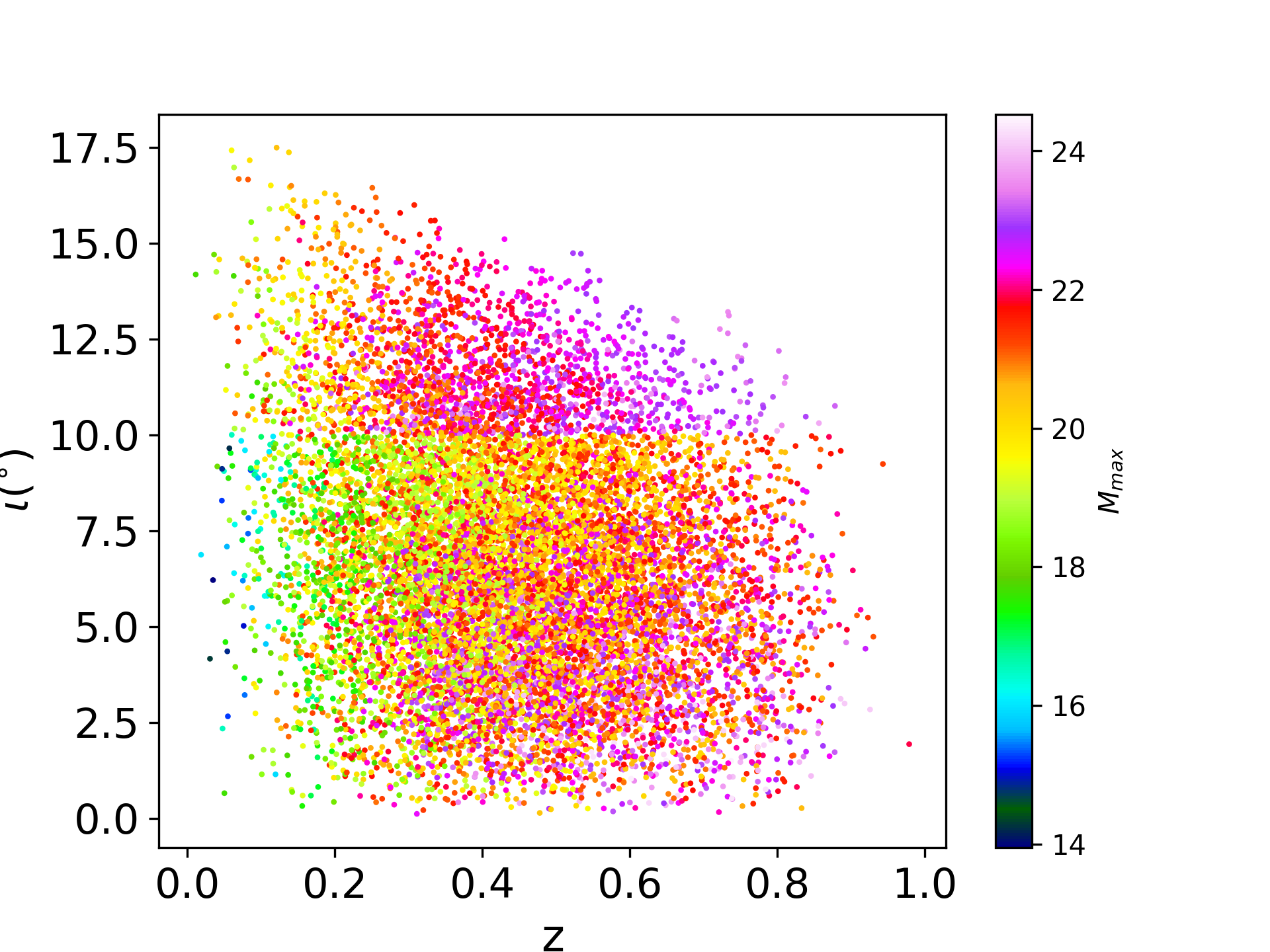}
}
\hspace{2pt}
\subfigure[CE, $n_{0}=0.1\ \mathrm{cm}^{-3}$]{
	\includegraphics[width=8cm]{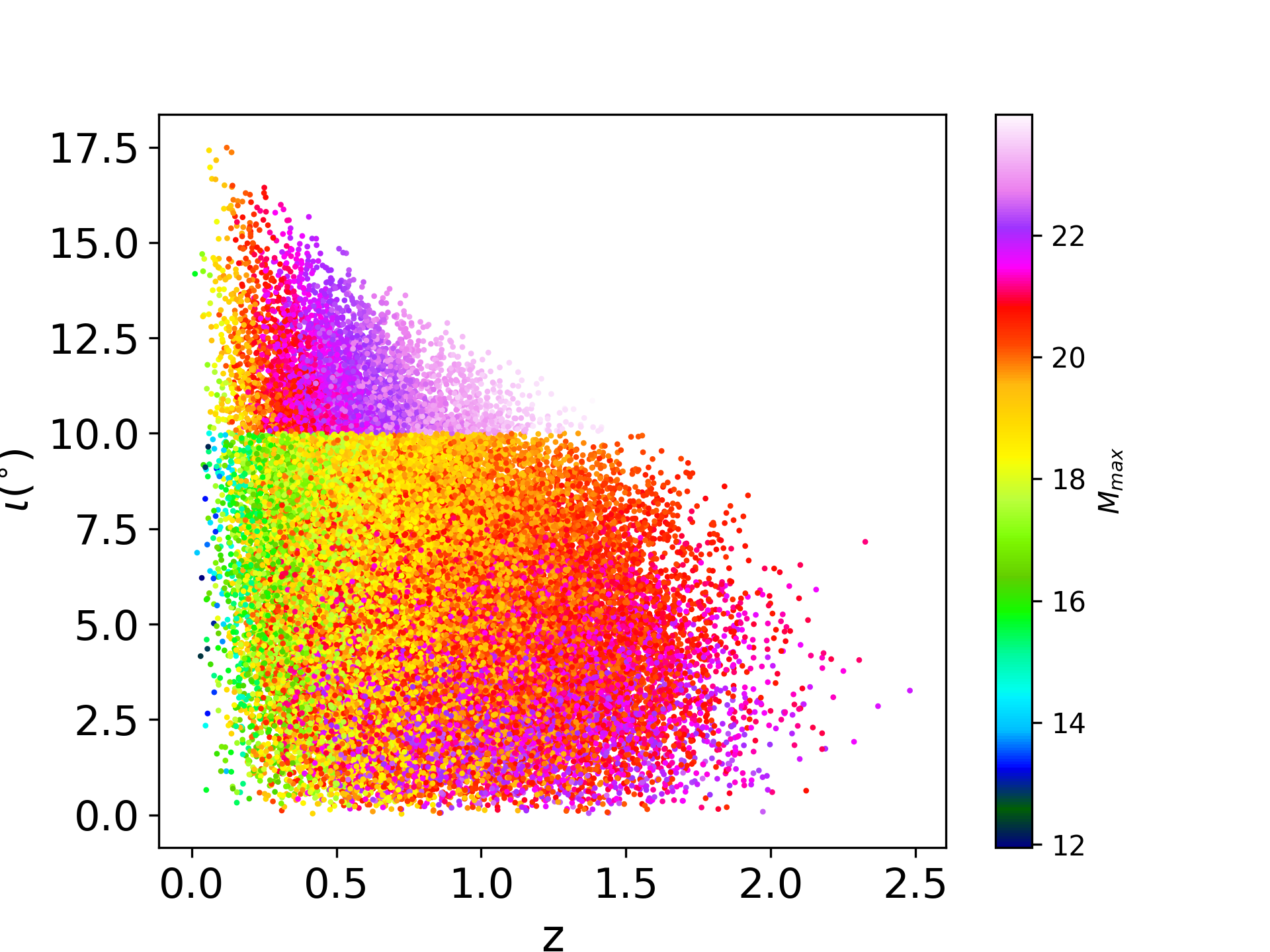}
}
\hspace{2pt}
\subfigure[CE, $n_{0}=0.01\ \mathrm{cm}^{-3}$]{
	\includegraphics[width=8cm]{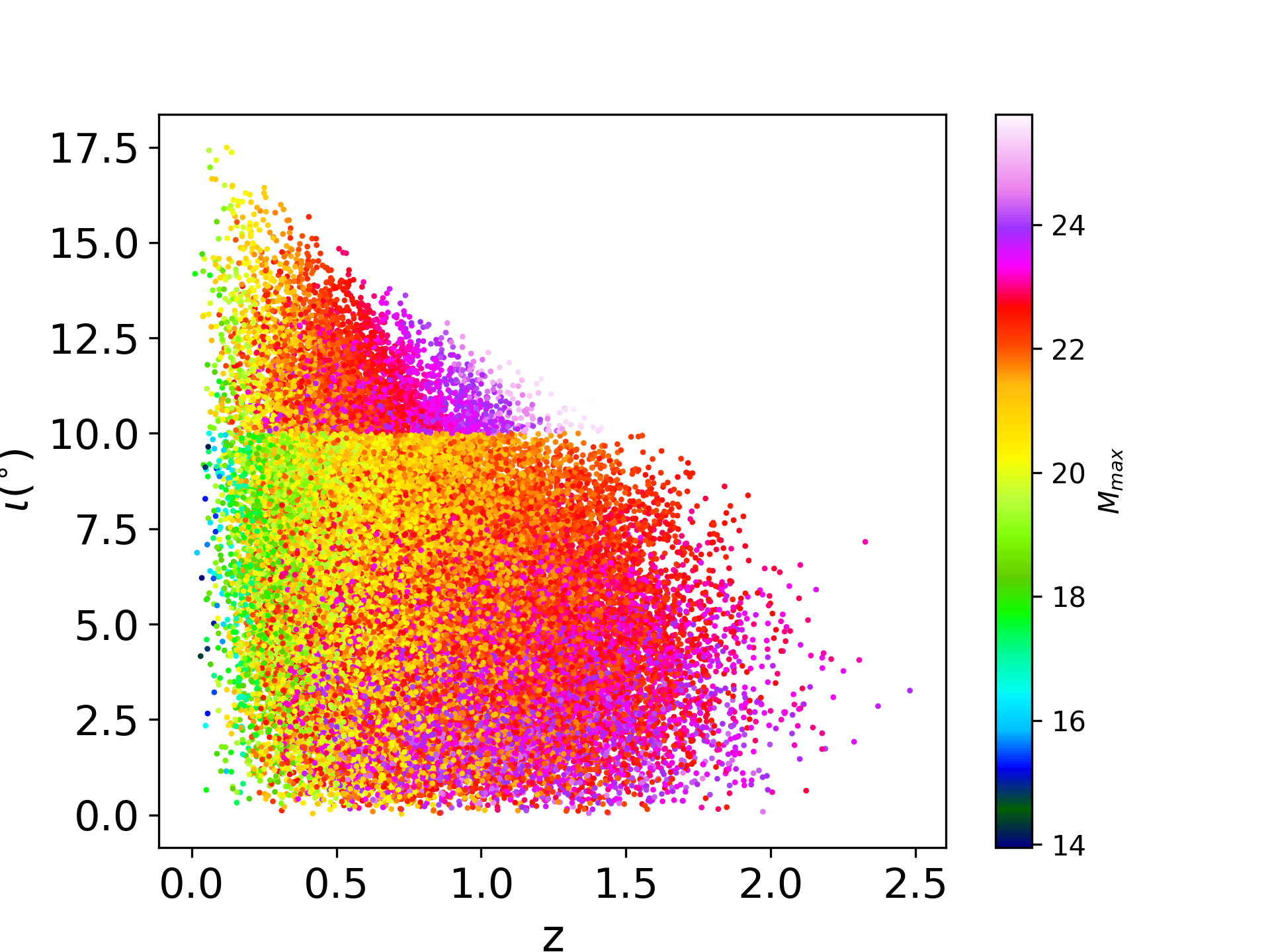}
}
\hspace{2pt}
\caption{The same with Fig. \ref{flux_scatter}, but with a different number density of the ISM. The left two panels show the results with $n_{0}=0.1\ \mathrm{cm}^{-3}$ and the right panels show the result with $n_{0}=0.01\ \mathrm{cm}^{-3}$.}
\label{flux_n0}
\end{figure*}

 
In Fig. \ref{flux_n0}, we set the number density of ISM as $n_{0}=0.1\ \mathrm{cm}^{-3}$ and $n_{0}=0.01\ \mathrm{cm}^{-3}$ as a comparison, respectively. GRB surrounded by thinner ISM usually has a weaker afterglow. In the case of $n_{0}=0.01\ \mathrm{cm}^{-3}$, a small fraction of afterglows are fainter than the design depth of LSST and WFST. In that case, the measurement of GRBs' redshift seems difficult and further discussion is needed.  \par

\begin{table*}
\begin{center}
\begin{tabular}{|c|c|c|c|c|c|}
\hline
&Swift-BAT&SVOM-ECLAIRS&GECAM&Fermi-GBM&EP\\
\hline
LHV&0.042-0.425 &0.072-0.731&0.278-2.820&0.198-2.001&0.029-0.297\\
\hline
LHVIK&0.084-0.856&0.146-1.474&0.553-5.598&0.394-3.985&0.058-0.593\\
\hline
LHV A+&0.217-2.200&0.374-3.789&1.370-13.870&0.962-9.741&0.148-1.504\\
\hline
LHVIK A+&0.445-4.505&0.766-7.757&2.743-27.769&1.907-19.305&0.301-3.046\\
\hline
ET&17.0-172.0&29.2-296.1&80.6-815.6&49.9-504.9&10.7-108.5\\
\hline
CE&98.1-993.0&168.9-1710.0&342.1-3463.5&188.4-1907.4&58.1-587.9\\
\hline
\end{tabular}
\end{center}
\caption{The multi-messenger observation rates (with the unit of year$^{-1}$) for BNS mergers with different $\gamma$-ray detectors and GW detectors.}
\label{table2}
\end{table*}


\subsection{Determination of Dark Energy}
\label{dark_energy}
In the last two decades, the accelerated expansion of the universe has been confirmed by a series of observations \citep{kowalski2008, hicken2009, komatsu2009, eisenstein2005, percival2007, percival2010, schrabback2010, kilbinger2009}. Various models of dark energy have been proposed to explain it (see \cite{copeland2006} for a review). In order to understand the physical properties of dark energy, it is important to measure its equation of state (EOS). In \cite{sath2010} and \cite{zhao2011}, the new method of measuring the dark energy EOS using GW standard sirens was proposed. \cite{zhao2018} compared the results of GW detector networks with 1000 face-on BNS sources, with baryon acoustic oscillations (BAO) method, and Type Ia supernovas (SNIa) method, and found that by the 3G GW detector networks, the constraints of dark energy parameters are similar with the SNIa and BAO methods. In this subsection, we will discuss these constraints with a quantitatively distributed BNS sample.\par
Similar to \cite{zhao2011}, we adopt a phenomenological form for EOS parameter $w$ as a function of redshift \citep{chevallier2001}
\be
	w(z)\equiv\frac{p_{\mathrm{de}}}{\rho_{\mathrm{de}}}=w_{0}+w_{a}\frac{z}{1+z},
\ee
where $p_{\mathrm{de}}$ is the pressure of dark energy and $\rho_{\mathrm{de}}$ is the energy density. The parameter $w_{0}$ represent the present EOS and $w_{a}$ represent the evolution with redshift. In the $\Lambda$CDM model, the relation of $d_{L}-z$, is determined by $(H_{0},\ \Omega_{m},\ \Omega_{k},\ w_{0},\ w_{a})$ together \citep{weinberg2008}. So if the $d_{L}-z$ relation can be measured from a series of multi-messenger observations, the set of these 5 parameters would have chance to be constrain. However, due to the strong degeneracy between the background parameters $(H_{0},\ \Omega_{m},\ \Omega_{k})$ and the dark energy parameter $(w_{0},\ w_{a})$, the constraints of full parameters set seems unrealistic \citep{zhao2011}. The same problem also happens in other methods for dark energy detection (e.g., SNIa and BAO methods) \citep{taskforce, zhao2011}. A general way to break this degeneracy is to combine the result with the CMB data, which are sensitive to the background parameters $(H_{0},\ \Omega_{m},\ \Omega_{k})$, and provide the necessary complement to the GW data. It has also been discovered in \citep{zhao2011} that taking the Planck CMB observation as a prior is nearly equivalent to treating the parameters $(H_{0},\ \Omega_{m},\ \Omega_{k})$ as known in the data analysis. Thus, similar to \cite{zhao2018}, \cite{zhaoLi2018} and \cite{yan2020}, in this article we use the GW data to constrain the dark energy parameters $(w_{0},\ w_{a})$ only.\par

\begin{figure*}[htbp]
\centering
\subfigure[ET]{
	\includegraphics[width=8cm]{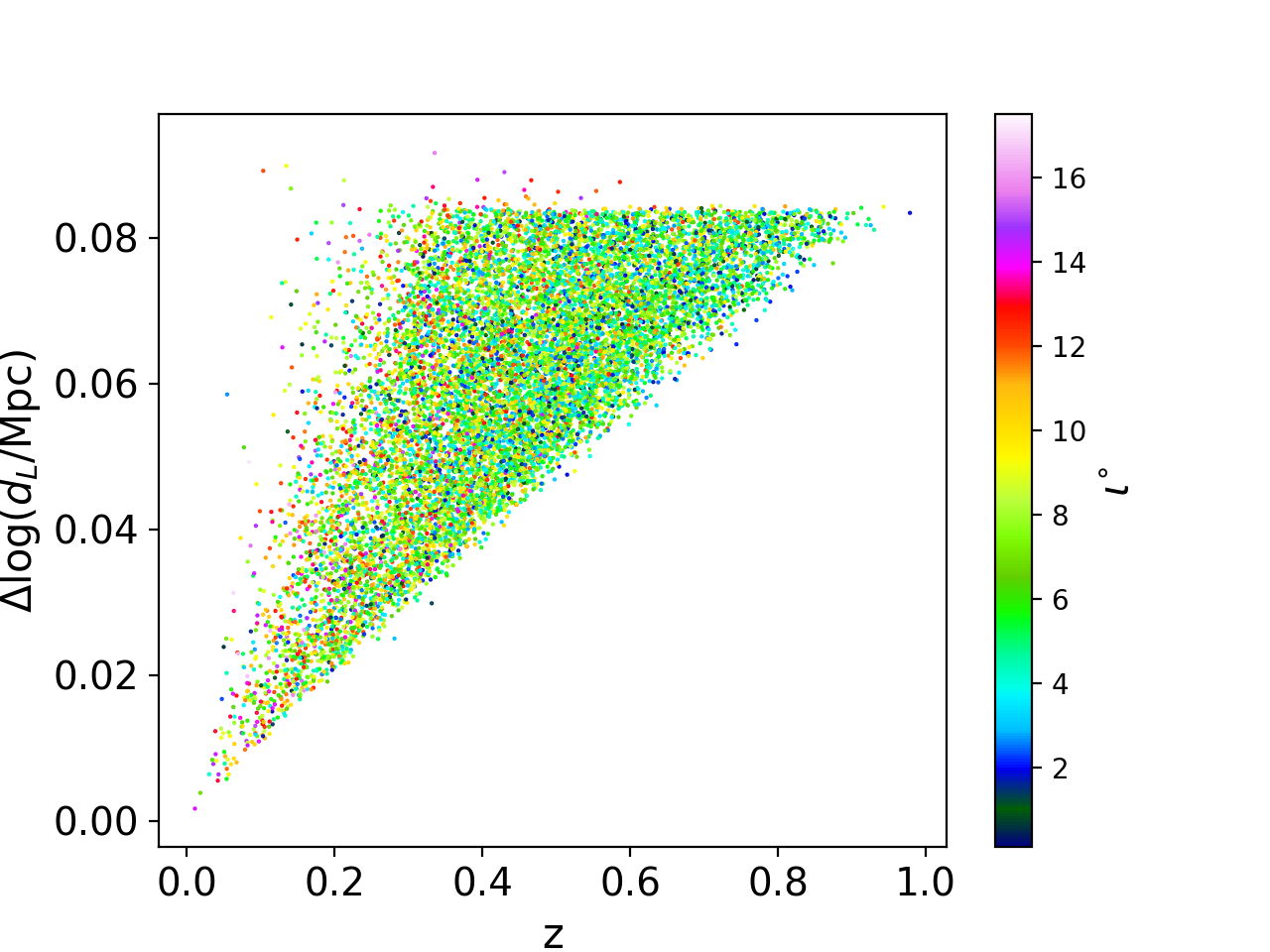}
}
\hspace{2pt}
\subfigure[CE]{
	\includegraphics[width=8cm]{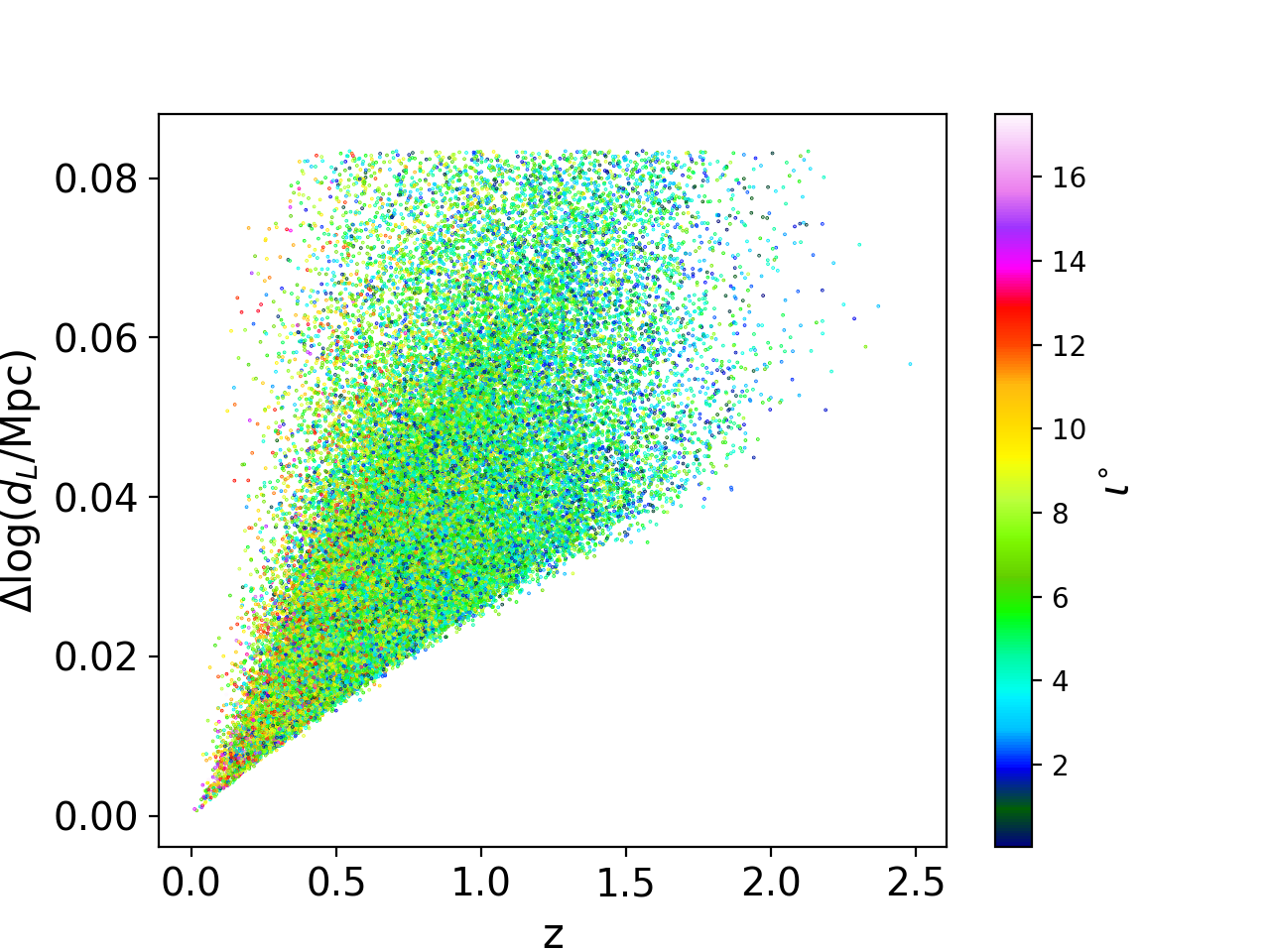}
}
\caption{The distribution of samples' $\Delta d_{L}$ and redshift, which can be triggered by 3G interferometers and GECAM. The left, right panels are the results with ET and CE, respectively. The color bar shows the samples' inclination angle.}
\label{DdL_1}
\end{figure*}

To estimate the errors of $(w_{0},\ w_{a})$, we consider a Fisher matrix $F_{ij}$ for a collection of $N$ BNS mergers that follows our sample distribution, which is given by \citep{zhao2011}
\be
	F_{ij}=\sum_{k=1}^{N}\frac{\partial\ln d_{L}(z_{k})}{\partial p_{i}}\frac{\partial\ln d_{L}(z_{k})}{\partial p_{j}}\frac{1}{(\delta d_{L}/d_{L}(\hat{\gamma}_{k},\ z_{k}))^2},
\ee
where $i$ and $j$ run from 1 to 2, denoting the free parameter $w_{0}$ and $w_{a}$, and $\hat{\gamma}$ represents the angle $(\alpha,\ \delta,\ \iota,\ \psi)$. The distance error $\delta d_{L}/d_{L}$ have two components: the instrument error $\Delta d_{L}/d_{L}$, and the error caused by the effects of weak lensing. We denote this error as $\tilde{\Delta}d_{L}/d_{L}$ and assume it satisfies $\tilde{\Delta}d_{L}/d_{L}=0.05z$ \citep{sath2010, zhao2011, zhao2018, Wang2019}. So the total distance error could be estimated from $\delta d_{L}/d_{L}=\sqrt{\left(\Delta d_{L}/d_{L}\right)^2+\left(\tilde{\Delta}d_{L}/d_{L}\right)^2}$. Similar to \cite{sathya2019} and \cite{Zhao&Santos2019}, in order to evaluate $\Delta d_L$ for each event, we assume the parameters $(\alpha,\ \delta,\ \iota)$ are constrained well by $\gamma$-ray observation and consider a Fisher matrix of six parameters, $(\psi,\ M,\ \eta,\ t_{c},\ \phi_{c},\ \log d_{L})$. This scenario is possible as we have already have seen in the case of GW170817. The sky position of GW170817 was constrained by finding the host galaxy NGC 4993 \citep{27} whereas the inclination angle was constrained from the X-ray
and ultraviolet observations \citep{28}. In Fig. \ref{DdL_1} we use the case of GECAM as a representative case and show the distribution of samples instrument errors with two single 3G GW detectors, ET and CE. There is an obvious cutoff around $\Delta\log\ d_{L}\sim0.08$, since the instrument errors have a strong correlation with SNR and we choose SNR$=12$ as the threshold. \par
\begin{figure*}[htbp]
\centering
\subfigure[ET and $R_{\mathrm{BNSmergers},0}=80\ \mathrm{Gpc}^{-3}\mathrm{yr}^{-1}$.]{
	\includegraphics[width=6.5cm]{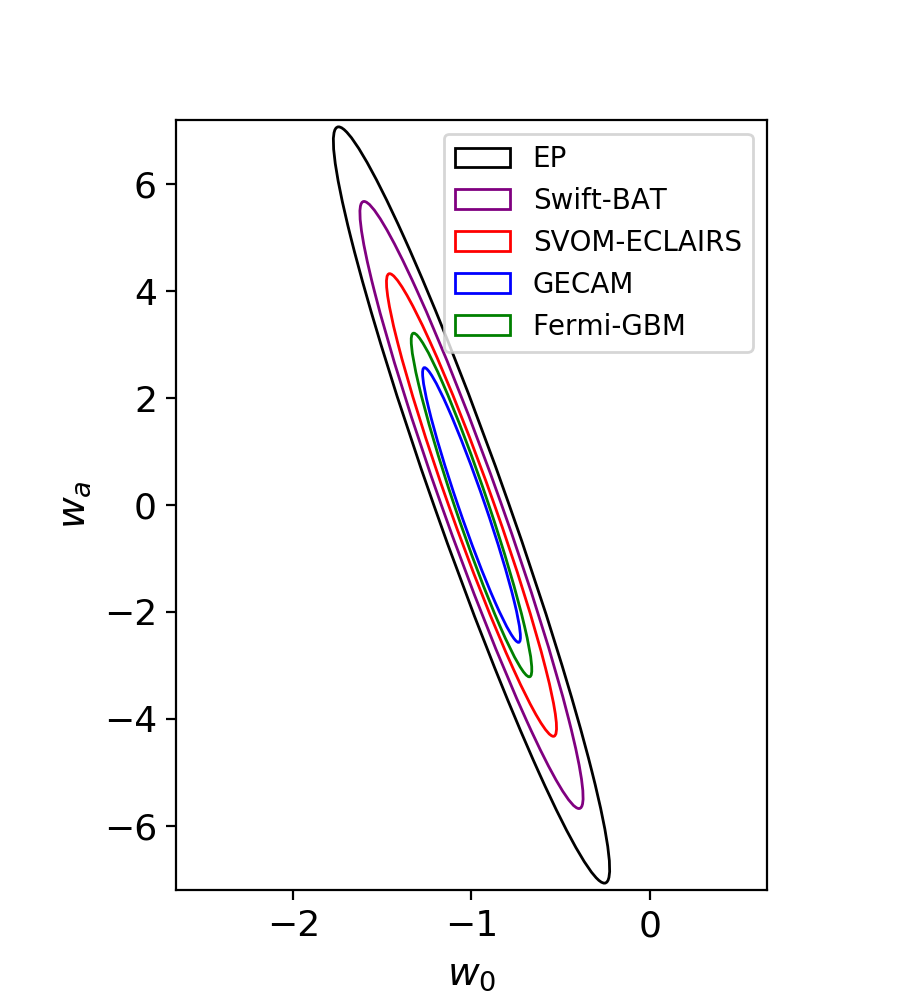}
}
\hspace{2pt}
\subfigure[ET and $R_{\mathrm{BNSmergers},0}=810\ \mathrm{Gpc}^{-3}\mathrm{yr}^{-1}$.]{
	\includegraphics[width=6.5cm]{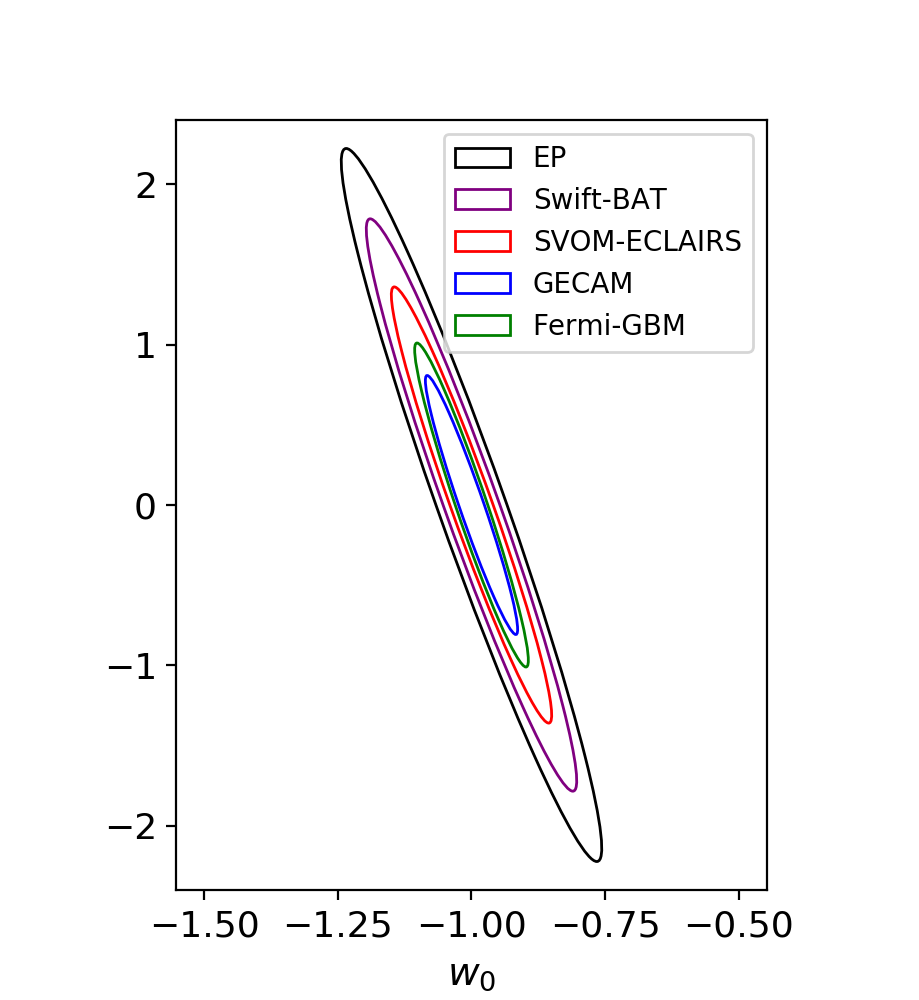}
}
\hspace{2pt}
\subfigure[CE and $R_{\mathrm{BNSmergers},0}=80\ \mathrm{Gpc}^{-3}\mathrm{yr}^{-1}$.]{
	\includegraphics[width=6.5cm]{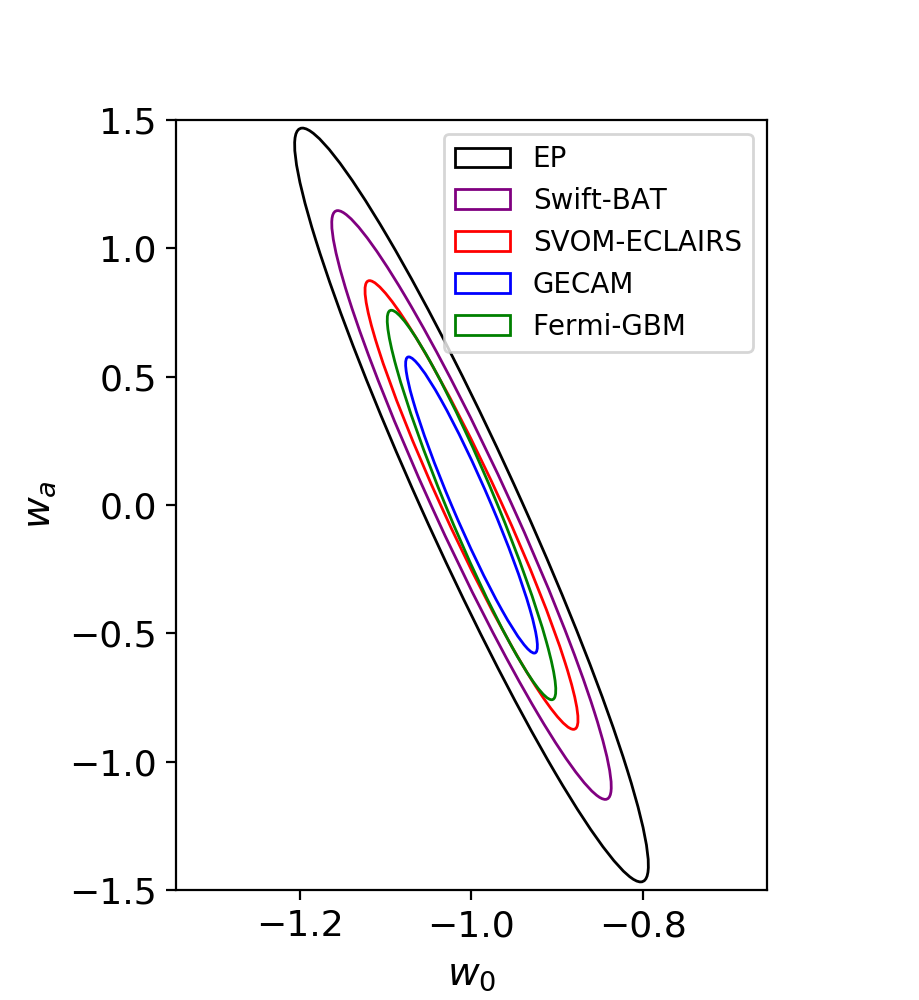}
}
\hspace{2pt}
\subfigure[CE and $R_{\mathrm{BNSmergers},0}=810\ \mathrm{Gpc}^{-3}\mathrm{yr}^{-1}$.]{
	\includegraphics[width=6.5cm]{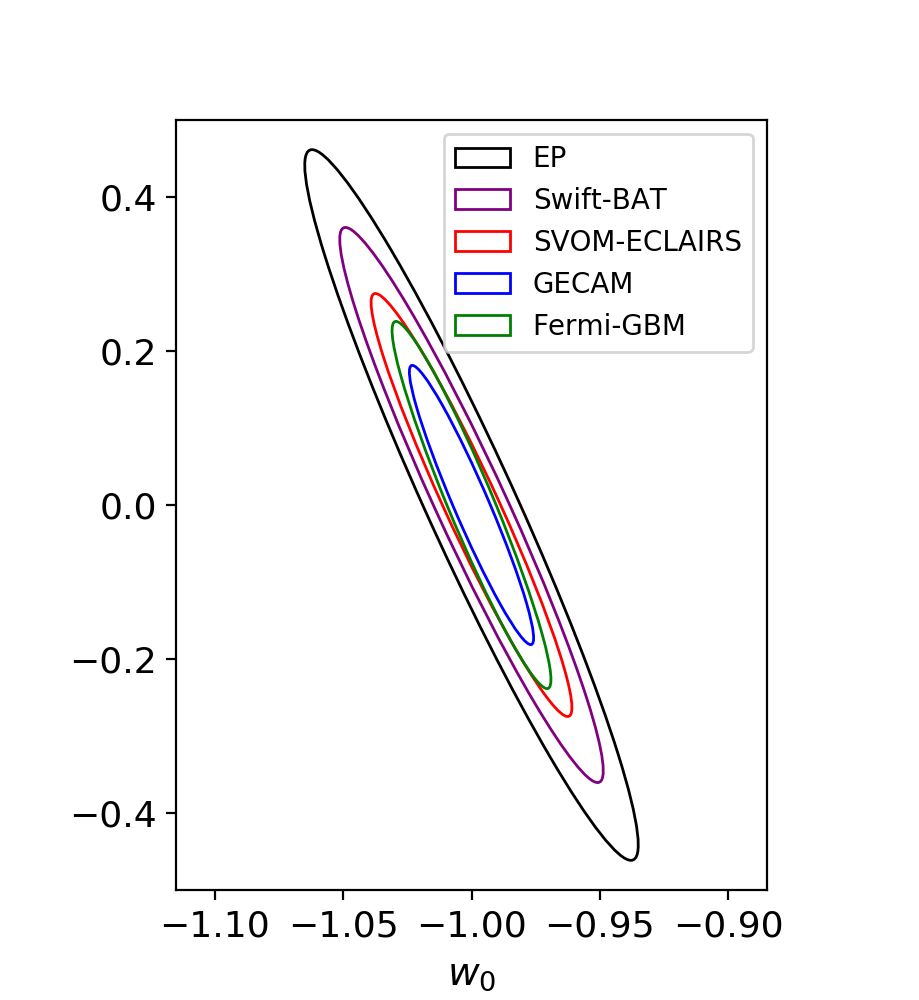}
}
\caption{The two-dimensional uncertainty contours of the dark energy parameters $w_{0}$ and $w_{a}$ for one year's multi-messenger observation of ET, CE and five $\gamma$-ray detectors with a 68.3\% ($1-\sigma$) confidence interval. The left panels are the results with $R_{\mathrm{BNSmergers},0}=80\ \mathrm{Gpc}^{-3}\mathrm{yr}^{-1}$ and the right panels are the results with $R_{\mathrm{BNSmergers},0}=810\ \mathrm{Gpc}^{-3}\mathrm{yr}^{-1}$.}
\vspace{5em}
\label{w0wa_ET}
\end{figure*}

The uncertainties of cosmological parameters are given by $\Delta w_{0}=(F^{-1})_{11}$ and $\Delta w_{a}=(F^{-1})_{22}$. If the statistical errors are dominant, the uncertainties of $w_{0}$ and $w_{a}$ are proportional to $1/\sqrt{N}$. In Table \ref{table5}, we list the 68.3\% (1-$\sigma$) uncertainties of $w_{0}$ and $w_{a}$ marginalized over other parameter with one year's multi-messenger observation. Also, to estimate the goodness of parameter constraints, we list the figure of merit (FoM) \citep{taskforce} in Table \ref{table5}, which is written as
\be
    \mathrm{FoM}=\left[\mathrm{Det}\ C(w_{0},w_{a})\right ]^{-1/2},
\label{FoM}
\ee
where $C(w_{0},w_{a})$ is the covariance matrix of $w_{0}$ and $w_{a}$. With the large observation area, the results of GECAM and Fermi-GBM are much better than other $\gamma$-ray detectors. For GECAM and CE, we obtain $\Delta w_{0}=0.016-0.051$, $\Delta w_{a}=0.119-0.380$ and $\mathrm{FoM}=171-1732$, for Fermi-GBM, the results are $\Delta w_{0}=0.020-0.065$, $\Delta w_{a}=0.157-0.499$ and $\mathrm{FoM}=100-1009$. These results are comparable with the detection abilities of SNIa  and BAO methods in the long-term (Stage IV) projects, which are $\mathrm{FoM}\sim 300$ for SNIa method and  $\mathrm{FoM}\sim 100$ for BAO method \citep{taskforce,zhao2011}. We can also see that the ET has poor ability of constraining $w_{0}$ and $w_{a}$ due to its low observation limit. In Fig. \ref{w0wa_ET}, the corresponding two-dimensional 1-$\sigma$ uncertainty contours of these multi-messenger observations with ET and CE  are plotted, and in each figure, we compare the results with four different $\gamma$-ray detectors. For the case of GECAM, the constraints of dark energy parameters are significantly more stringent than other three $\gamma$-ray detectors due to its large observation areas. The results of CEET and CE2ET are not showed here because their shapes are very similar with the results with CE. Our results roughly match the results of \cite{zhao2018}, but using a much more quantitative BNS sample. \par

\begin{table*}
\begin{center}
\begin{tabular}{|c|c|c|c|c|c|c|}
\hline
\multicolumn{2}{|c|}{}&Swift-BAT&SVOM-ECLAIRS&GECAM&Fermi-GBM&EP\\
\hline
\multirow{4}{*}{$\Delta w_{0}$}
&ET&0.129-0.412&0.099-0.314&0.057-0.181&0.070-0.222&0.160-0.510\\
\cline{2-7}
&CE&0.034-0.107&0.026-0.082&0.016-0.051&0.020-0.065&0.043-0.136\\
\cline{2-7}
&CEET&0.032-0.104&0.025-0.079&0.015-0.049&0.020-0.062&0.041-0.131\\
\cline{2-7}
&CE2ET&0.028-0.090&0.022-0.069&0.013-0.042&0.017-0.054&0.036-0.114\\
\hline
\multicolumn{7}{|c|}{}\\
\hline
\multirow{4}{*}{$\Delta w_{a}$}
&ET&1.173-3.734&0.894-2.846&0.531-1.690&0.664-2.114&1.462-4.652\\ 
\cline{2-7}
&CE&0.237-0.754&0.181-0.575&0.119-0.380&0.157-0.499&0.303-0.966\\
\cline{2-7}
&CEET&0.228-0.727&0.174-0.554&0.115-0.366&0.151-0.481&0.293-0.931\\
\cline{2-7}
&CE2ET&0.198-0.630&0.151-0.480&0.100-0.318&0.132-0.419&0.254-0.808\\
\hline
\multicolumn{7}{|c|}{}\\
\hline
\multirow{4}{*}{FoM}
&ET&2.4-24.4&4.1-42.0&11.7-118.8&7.4-75.0&1.5-15.6\\ 
\cline{2-7}
&CE&42.5-430.5&73.2-741.3&171.1-1732.0&99.7-1009.1&26.0-263.0\\
\cline{2-7}
&CEET&45.7-462.9&78.7-797.1&183.7-1860.3&107.0-1083.8&27.2-282.7\\
\cline{2-7}
&CE2ET&60.4-611.4&104.0-1052.9&241.7-2447.6&140.6-1423.9&36.8-372.9\\
\hline
\end{tabular}
\end{center}
\caption{The $1-\sigma$ constraints of $\Delta w_{0}$ and $\Delta w_{a}$ by one year's joint observations with different GW interferometers and $\gamma$-ray detectors. The FoM in each case is also showed.}
\label{table5}
\end{table*}


\section{Effect of alternative models for EM counterparts}
\label{discussion}
In previous discussions, we assume all BNS mergers have a Gaussian-shaped jet profile in the form of Eq. (\ref{jet}). In \cite{abbott2017d}, other two jet profiles are used to explain the observed properties of GRB 170817A, uniform top-hat jets and 'cocoon' emission model \citep{lazzati2017}. However, the compact radio emission observations favor the structured jet profile \citep{ghirlanda2019}. Therefore, in this article, we will not consider these two models. Meanwhile, to combine GRB 170817A and other SGRBs, \cite{tan2018} produce a two-Gaussian profile, 
\be
    L_{\mathrm{iso}}(\iota)=L_{\mathrm{on}}\left [ \exp\left ( -\frac{\iota^{2}}{2\theta_{\mathrm{in}}^{2}}\right )+\mathcal{C}\exp\left (-\frac{\iota^{2}}{2\theta_{\mathrm{out}}^{2}}\right )\right ],
\label{two_gaussian}
\ee
where $L_{\mathrm{on}}$, $\theta_\mathrm{in}$, $\theta_{out}$, $\mathcal{C}$ are four free parameters, with $\theta_{\mathrm{out}}=10\theta_{\mathrm{in}}$ and $\mathcal{C}\ll1$. We choose the $\theta_{\mathrm{in}}=2^{\circ}$ Model as example. The multi-messenger observation rates with this model are showed in Table \ref{table6}. For 2G GW network, the rates with Swift-BAT and SVOM are a few times larger than the case of Gaussian jet profile, since more GRBs in this model have peak energy in Swift-BAT and SVOM's flux band. Other rates are roughly same. For 3G interferometers, the rates are about half compared with Table \ref{table2}, since two-Gaussian profile is narrower around $\iota=0^{\circ}$.\par
\begin{figure}
    \centering
    \includegraphics[width=8cm]{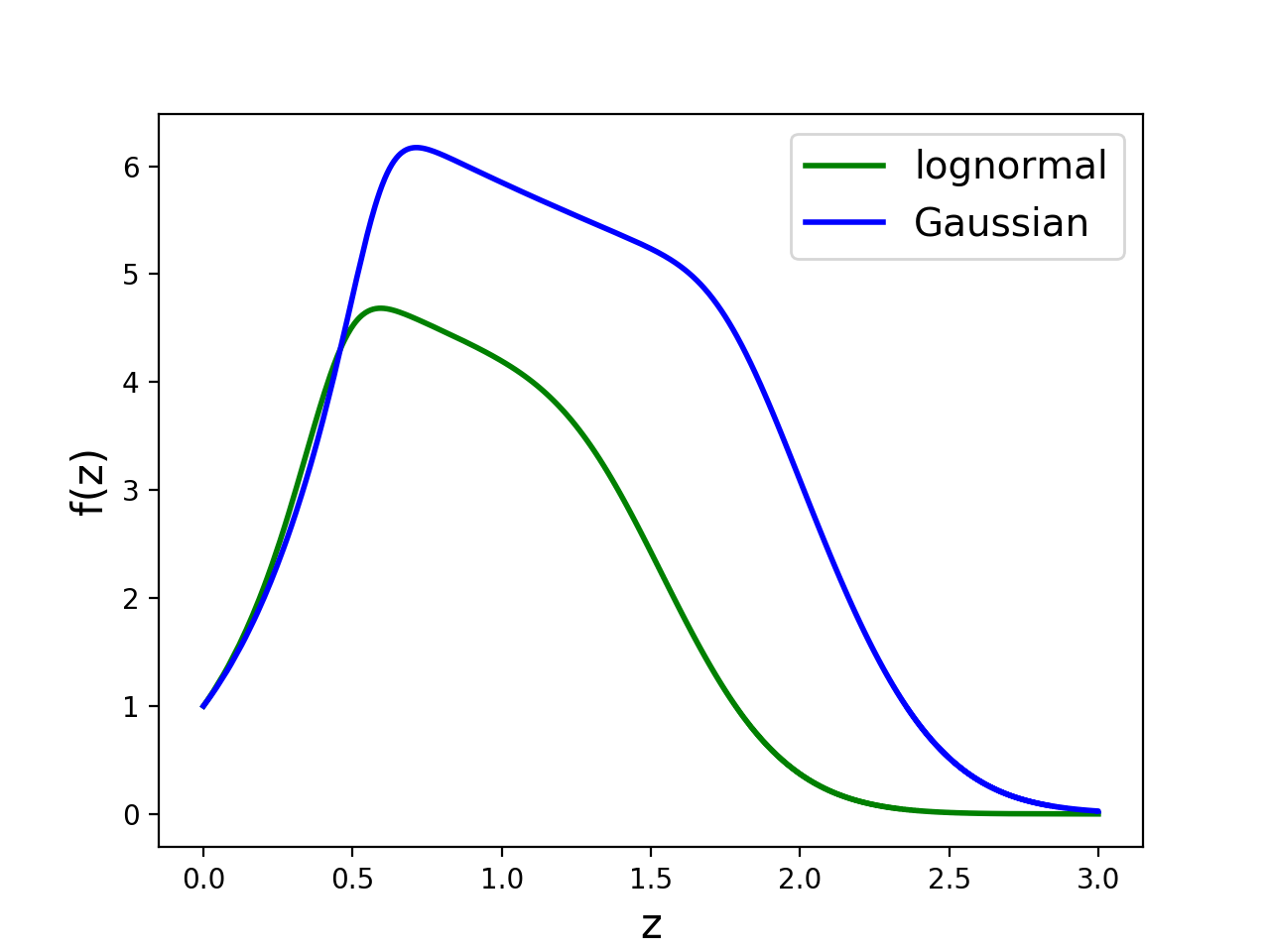}
    \caption{The redshift distribution of BNS merger samples. The green and blue lines represent the log-normal and Gaussian time delay models, respectively.}
    \label{fig_fz}
\end{figure}

In Sec. \ref{samples}, we use the log-normal distribution time delay model. In \cite{sun2015}, the authors investigated it with other two models, i. e. Gaussian delay model \citep{vir2011} and power-law delay model \citep{sun2015}. The power-law delay model is not supported by the SGRB data \citep{vir2011}. Therefore, we concentrate on the difference between Gaussian and log-normal delay models here. The $f(z)$ mentioned in Eq. (\ref{fz}) with these two models are painted in Fig. \ref{fig_fz} and the observation rates with Gaussian delay model are showed in Table \ref{table7}. While $z<0.5$, the distributions are almost same. So for 2G GW networks, the multi-messenger observation rates are also almost same. However, while for high redshift, the Gaussian delay model predicts a larger number of BNS samples, so the rates in the case of 3G interferometers, especially for CE, are lager than log-normal delay model.\par
\begin{figure*}[htbp]
\centering
\subfigure[2G]{
	\includegraphics[width=8cm]{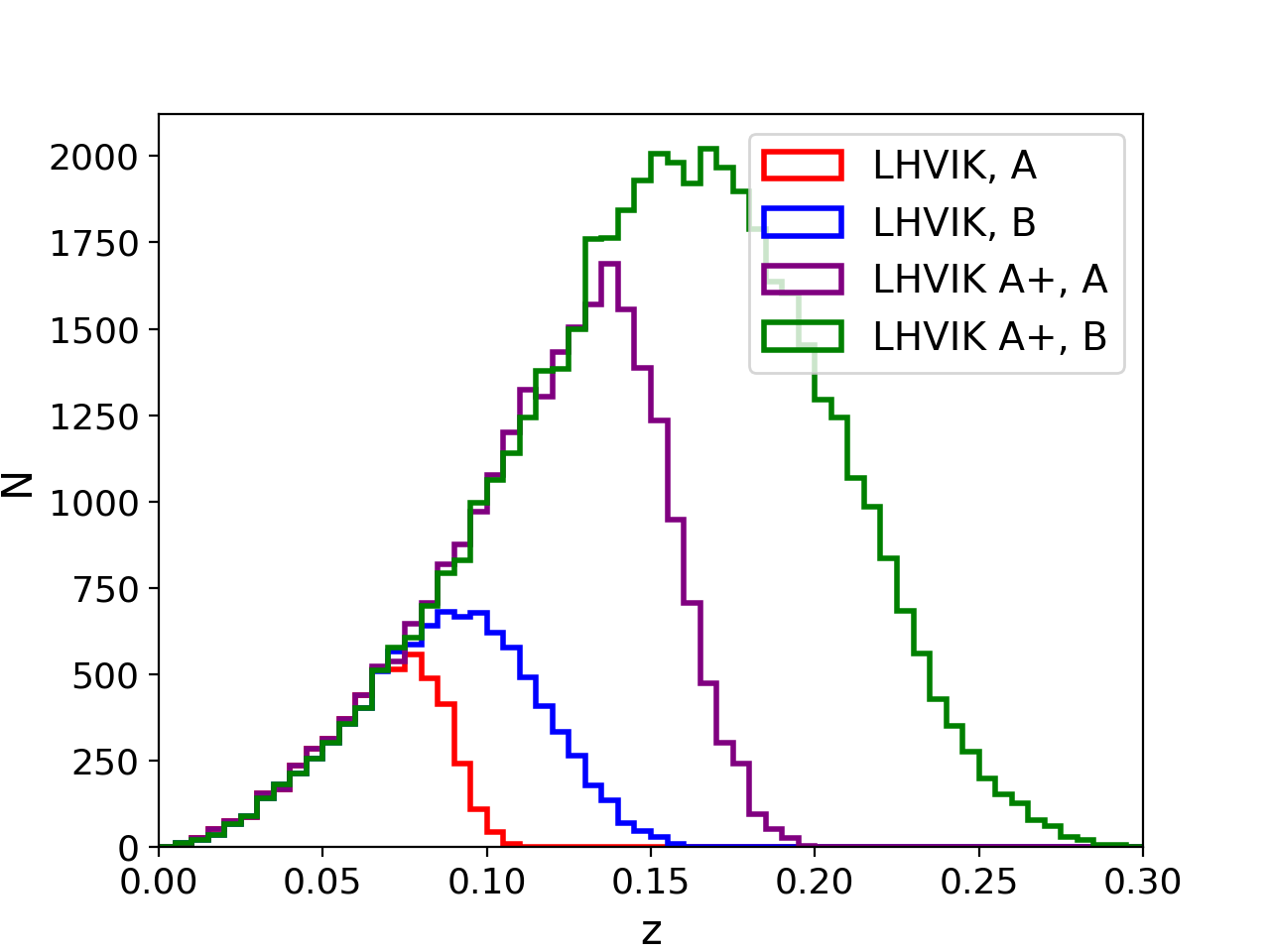}
}
\hspace{2pt}
\subfigure[3G]{
	\includegraphics[width=8cm]{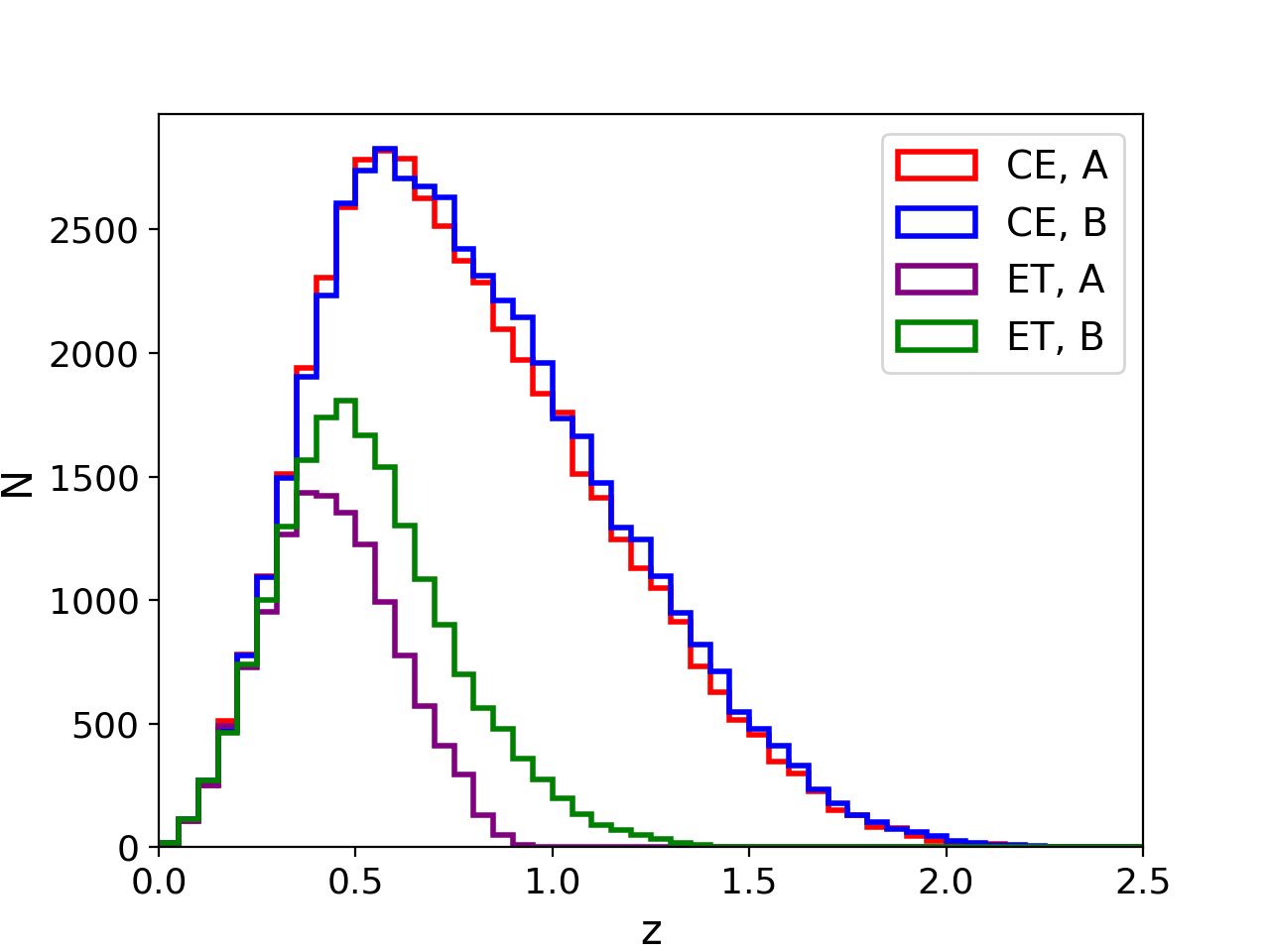}
}
\caption{The comparison of two distributions of BNS mass. Model A represents the normal distribution mentioned in Sec. \ref{samples} and model B represents the flat mass distribution mentioned in \cite{abbott2020e}.}
\label{pdf_z_compare}
\end{figure*}

Another issue that needs to be discussed is, GW190425 \citep{abbott2020a}, a compact binary coalescence with total mass $\sim3.4\ M_{\odot}$ does not fit in with the BNS mass distribution assumed in our work. In order to cover the GW190425-type systems, we follow the same setting as \cite{abbott2020e}, assume a uniform mass distribution between $1\ M_{\odot}$ and $2.5\ M_{\odot}$. In Fig. \ref{pdf_z_compare}, we compare the redshift distributions of BNS samples which can be both triggered by GW interferometers and GECAM with two BNS mass distributions. Due to the heavy BNS mergers considered, for 2G GW interferometers and ET, more and further BNS mergers can be observed. However, for CE, since the maximum observation distance exceeds the $\gamma$-ray detectors' and the dominant constraints come from GRB observations, the results of two mass distributions are very similar. In Table \ref{table8}, the merger rates with other $\gamma$-ray detectors and GW interferometers are listed. \par
\begin{figure*}[htbp]
\centering
\subfigure[ET and $R_{\mathrm{BNSmergers},0}=80\ \mathrm{Gpc}^{-3}\mathrm{yr}^{-1}$.]{
	\includegraphics[width=6.5cm]{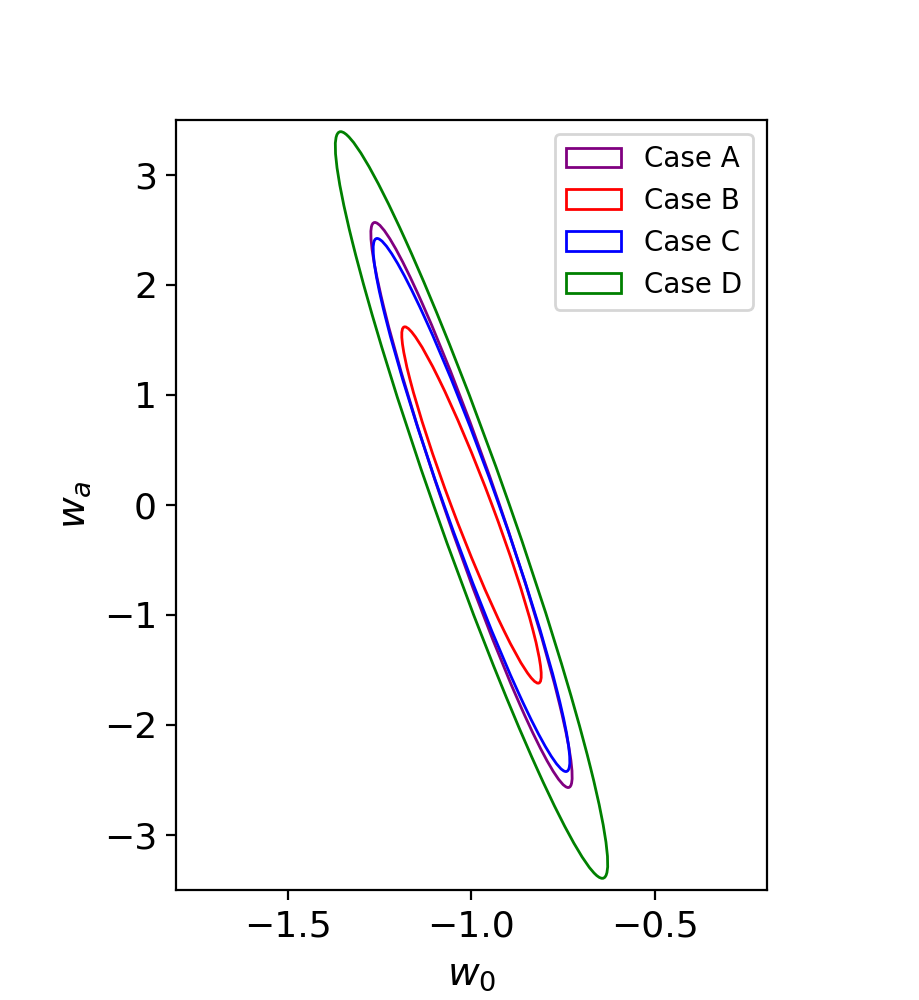}
}
\hspace{2pt}
\subfigure[ET and $R_{\mathrm{BNSmergers},0}=810\ \mathrm{Gpc}^{-3}\mathrm{yr}^{-1}$.]{
	\includegraphics[width=6.5cm]{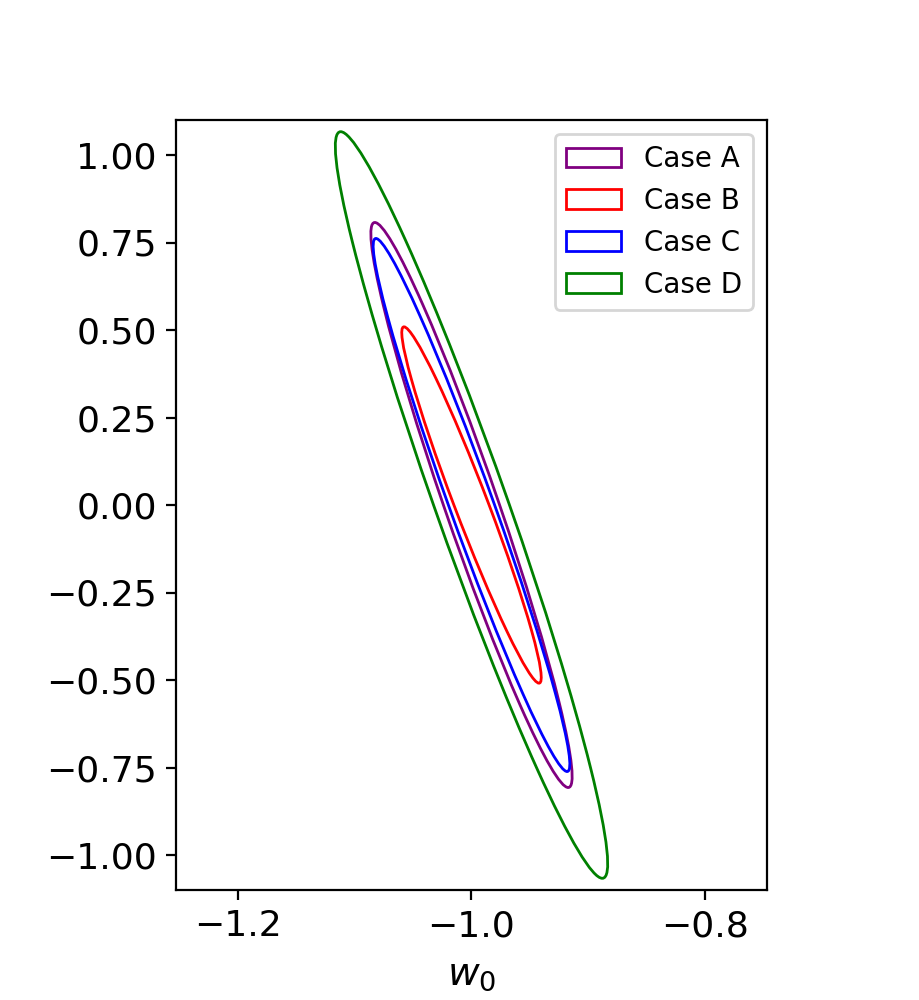}
}
\hspace{2pt}
\subfigure[CE and $R_{\mathrm{BNSmergers},0}=80\ \mathrm{Gpc}^{-3}\mathrm{yr}^{-1}$.]{
	\includegraphics[width=6.5cm]{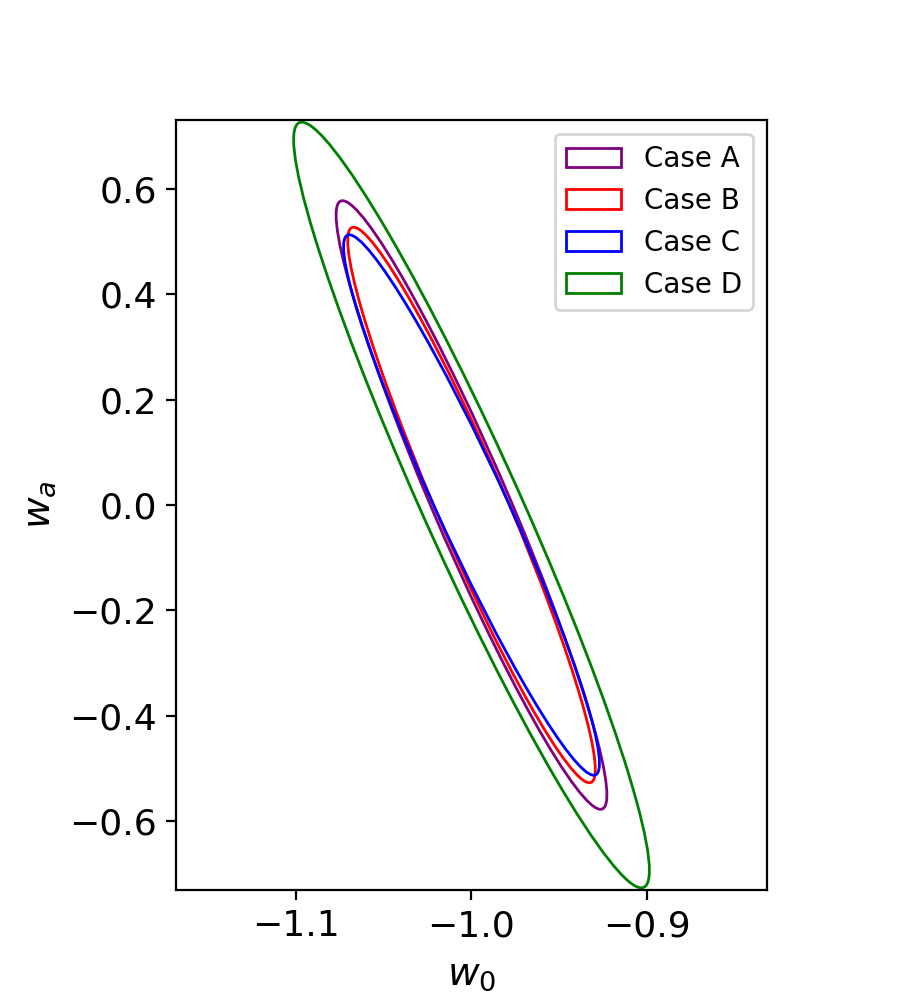}
}
\hspace{2pt}
\subfigure[CE and $R_{\mathrm{BNSmergers},0}=810\ \mathrm{Gpc}^{-3}\mathrm{yr}^{-1}$.]{
	\includegraphics[width=6.5cm]{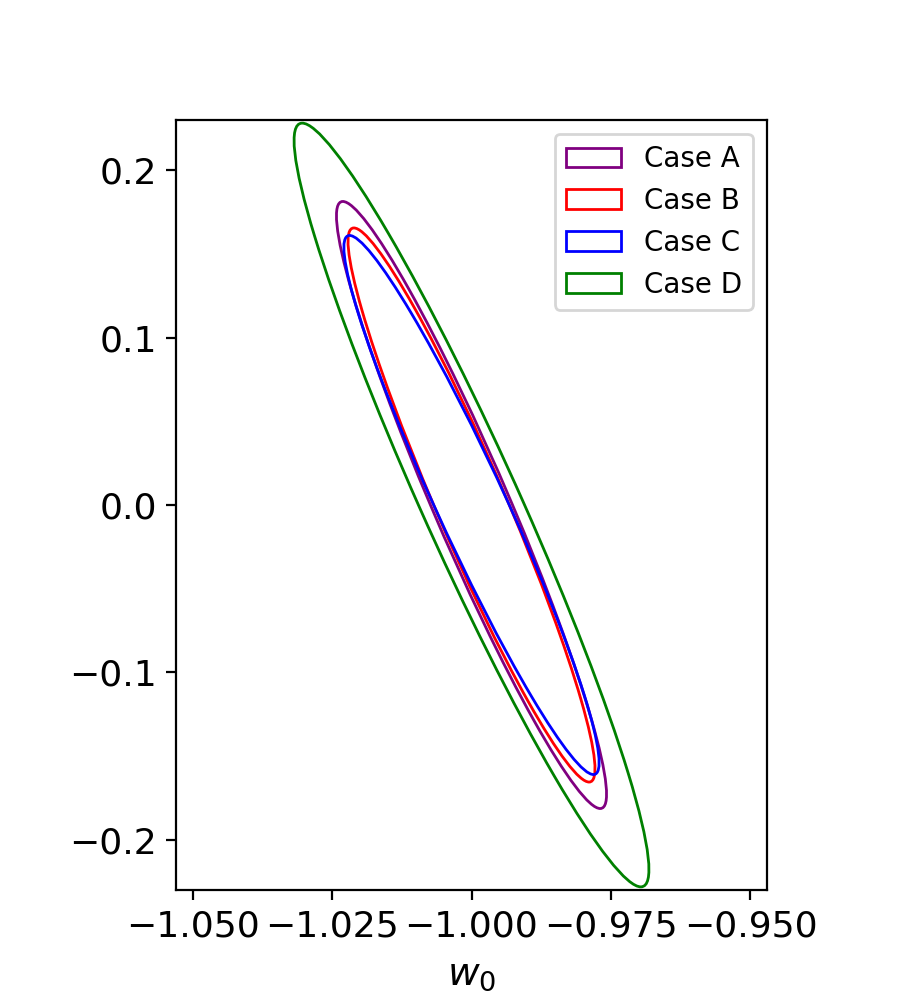}
}
\caption{The two-dimensional 1-$\sigma$ uncertainty contours of the dark energy parameters $w_{0}$ and $w_{a}$ with different BNS samples. Case A represents the results of delay time with lognormal distribution and a Gaussian BNS mass distribution, case B uses a flat mass distribution, case C uses Gaussian deley time model and case D uses two-Gaussian jet profile.}
\vspace{5em}
\label{w0wa_compare}
\end{figure*}

In this section, we discuss three different models as a comparison and predict the multi-messenger observation rates with each model. Following the method mentioned in Sec. \ref{dark_energy}, we calculate the constraints on $w_{0}$ and $w_{a}$ through multi-messenger observations with these three models. In Fig. \ref{w0wa_compare}, we compare the corresponding two-dimensional $1-\sigma$ uncertainty contours of $\Delta w_{0}$ and $\Delta w_{a}$ with four models. The contours with two-Gaussian jet profile are larger than others due to smaller observation rates. In the case of ET and flat mass distribution, the constraints are better. And in other cases, the results are roughly similar. \par


\begin{table*}
\begin{center}
\begin{tabular}{|c|c|c|c|c|c|}
\hline
&Swift-BAT&SVOM-ECLAIRS&GECAM&Fermi-GBM&EP\\
\hline
LHV&0.185-1.873 &0.318-3.226&0.384-3.884&0.183-1.850&0.055-0.561\\
\hline
LHVIK&0.315-3.194&0.543-5.500&0.580-5.868&0.290-2.936&0.075-0.760\\
\hline
LHV A+&0.529-5.356&0.911-9.222&1.007-10.198&0.563-5.702&0.117-1.186\\
\hline
LHVIK A+&0.791-8.011&1.362-13.795&1.680-17.015&1.018-10.305&0.187-1.890\\
\hline
ET&8.8-89.3&15.2-153.7&44.3-448.3&29.2-259.2&5.3-54.0\\
\hline
CE&45.1-457.0&77.7-786.9&202.1-2046.3&122.3-1228.7&29.5-298.4\\
\hline
\end{tabular}
\end{center}
\caption{The same with Table \ref{table2}, but with the jet profile in \cite{tan2018}.}
\label{table6}
\end{table*}


\begin{table*}
\begin{center}
\begin{tabular}{|c|c|c|c|c|c|}
\hline
&Swift-BAT&SVOM-ECLAIRS&GECAM&Fermi-GBM&EP\\
\hline
LHV&0.041-0.411 &0.070-0.708&0.265-2.681&0.189-1.916&0.028-0.287\\
\hline
LHVIK&0.082-0.829&0.141-1.427&0.528-5.347&0.376-3.810&0.057-0.577\\
\hline
LHV A+&0.208-2.112&0.359-3.636&1.308-13.239&0.922-9.332&0.142-1.444\\
\hline
LHVIK A+&0.432-4.373&0.744-7.531&2.660-26.938&1.856-18.797&0.293-2.970\\
\hline
ET&18.3-185.7&31.5-319.8&85.0-860.7&53.0-536.2&11.5-116.4\\
\hline
CE&130.2-1318.2&224.2-2268.9&423.4-4287.4&227.4-2303.0&75.9-768.9\\
\hline
\end{tabular}
\end{center}
\caption{The same with Table \ref{table2}, but with a Gaussian time delay model.}
\label{table7}
\end{table*}

\begin{table*}
\begin{center}
\begin{tabular}{|c|c|c|c|c|c|}
\hline
&Swift-BAT&SVOM-ECLAIRS&GECAM&Fermi-GBM&EP\\
\hline
LHV&0.079-0.800&0.136-1.378&0.506-5.129&0.360-3.648&0.054-0.551\\
\hline
LHVIK&0.163-1.655&0.281-2.850&1.033-10.461&0.730-7.392&0.112-1.132\\
\hline
LHV A+&0.421-4.264&0.725-7.343&2.563-25.944&1.780-18.021&0.284-2.878\\
\hline
LHVIK A+&0.881-8.917&1.516-15.354&5.260-53.257&3.612-36.572&0.590-5.978\\
\hline
ET&27.3-276.2&47.0-475.7&119.4-1209.0&71.6-724.6&16.9-171.4\\
\hline
CE&90.6-917.4&156.0-1579.7&318.2-3221.7&176.0-1781.6&53.8-545.0\\
\hline
\end{tabular}
\end{center}
\caption{The same with Table \ref{table2}, but with a flat mass distribution between $1\ M_{\odot}$ and $2.5\ M_{\odot}$.}
\label{table8}
\end{table*}




\section{Conclusions}
\label{conclusion}
The detection of GW170817 opens a door of multi-messenger observation. From the GW waveforms of these events, we could measure the sources' luminosity distances independently, without the comic distance ladder. If the redshifts of the sources could also be measured from other methods, this kind of GW events could be treated as the standard sirens to constrain cosmological parameters, such as Hubble constant, dark energy and so on. A direct way to measure the sources' redshifts is to find their host galaxies by the observation of their EM counterparts, such as GW170817. \par
In this paper, we discuss the multi-messenger observations of BNS mergers' GW and GRB signals. We find the network of the 2.5G network LHVIK A+ are expect to detect more than 3 multi-messenger signals per year with different $\gamma$-ray detectors, if the local merger rate is 810 Gpc$^{-3}$yr$^{-1}$. And the upper redshift limit of these BNS samples is about 0.2. {\color{black}For the LHVIK or LHV A+, the numbers are about several times smaller than LHVIK A+.} Most of the BNS mergers which could be triggered by both GW interferometers and $\gamma$-ray detectors have $\iota\leq22^{\circ}$, because of the limitation of the Gaussian jet profile. Only a few samples in very low redshift with inclination angles larger than 22$^{\circ}$ can be triggered by $\gamma$-ray detectors. For the 3G interferometer ET and CE, the number of the multi-messenger observation rates are tens to hundreds of times larger compared with LHVIK A+. The upper redshift limit for ET is $\sim1$ and for CE, the upper limit is $\sim2.5$. We also find for the samples with $z$ larger than 1.0, the requirements on the inclination angles become stricter, becoming less than $15^\circ$.  The results of 3G GW detectors are in line with the magnitude estimates of BNS event rates in \cite{sath2010} and \cite{zhao2011}. BNS mergers' inclinations that can be observed by both GW and $\gamma$-ray detectors are almost all less than $22^{\circ}$, which also match these works' estimate. 

Due to the high redshift upper limit of the multi-messenger observations with the 3G GW detectors, we discuss the implication for constraining dark energy parameters with BNS mergers. One year's observations of ET can constrain the EOS of cosmic dark energy with accuracies $\Delta w_{0}$ $\sim$ 0.06-0.18 and $\Delta w_{a}\sim0.5-1.7$ with GECAM. For CE, the results are much better, which are $\Delta w_{0}\sim0.02-0.05$ and $\Delta w_{a}\sim 0.1-0.4$. We also discuss the cases of 3G GW detector networks, CEET and CE2ET. The results of CEET are quite close to CE and the results of CE2ET have a $\sim$ 10-20\% progress. Therefore, the usage of 3G interferometers will provide an important method to help us understand the high-redshift Universe.\par

Note that, \cite{belgacem2019} also estimated the multi-messenger observation rates, where the authors used a Gaussian structured jet profile with specific parameters and selected observed GRBs through their peak flux. In this work, we make a much more comprehensive and integrated analysis on this issue. We discuss a more general structured jet profile and calculate the spectrum of each GRB sample from the Band function, and integrate it in the band of different GRB detectors, instead of peak flux. Because of these, our predictions are several times larger than the results in \cite{belgacem2019}. In order to study the uncertainties in BNS samples and GRB jets, we discuss several alternative models. The two-Gaussian jet profile, the Gaussian time delay model and a flat BNS mass distribution are discussed as a supplement. The two-Gaussian jet profile model brings a narrower jet and a softer spectrum. The results of 2G GW networks and Swift-BAT/SVOM are $\sim 4$ times larger and the rates with 3G GW detectors are about half. The Gaussian delay time model predicts a larger BNS merger rate in high redshift and the results of 3G interferometers are slightly larger. The flat BNS mass distribution takes heavy BNS mergers into account, and the rates with ET have a $\sim50\%$ increase. But for CE, due to the constraints from GRB observations, the rates are almost same. We also estimate the magnitudes of optical afterglows and find that, if the number density of ISM is $1\ \mathrm{cm}^{-3}$, due to the low inclination angle, all multi-messenger events have optical afterglows that reach the single-visit depth of LSST and/or WFST. \par


~

\acknowledgments

This work is supported by the National Natural Science Foundation of China under Grant Nos. 11903030, 11773028, 11603020, 11633001, 11653002, the Strategic Priority Program of the Chinese Academy of Sciences (Grant No. XDB 23040100) and the Fundamental Research Funds for the Central Universities under Grant Nos: WK2030000036 and WK3440000004.


\end{document}